\documentclass[preprint, 11pt]{imsart}
\usepackage[english]{babel}
\usepackage{amsmath,amssymb,amsthm}
\usepackage[latin9]{inputenc}
\usepackage[T1]{fontenc}
\usepackage{graphicx, xcolor,tikz,arrayjobx,trimspaces,subfig}
\usepackage[labelfont=bf]{caption}
\usepackage{enumerate,multirow}
\usepackage{soul}

\usepackage[mathlines]{lineno}             
\usepackage[round, sort, numbers, authoryear]{natbib}   
\usepackage[colorlinks=true,citecolor=blue,urlcolor=blue]{hyperref}                    

\usepackage{color}
\usepackage{marvosym}  

\usepackage{pgfplots}

\pgfplotsset{compat=1.11}
\usetikzlibrary{calc}
\usetikzlibrary{positioning}
\usetikzlibrary{decorations.pathreplacing}

\usepackage[toc, page]{appendix}

\usepackage{todonotes}   

\newtheorem{theo}{Theorem}[section]
\newtheorem{mydef}[theo]{Definition}
\newtheorem{mylemma}[theo]{Lemma}
\newtheorem{myprop}[theo]{Proposition}
\newtheorem{mycoro}[theo]{Corollary}
\theoremstyle{definition}
\newtheorem{example}{Example}[section]
\theoremstyle{remark}
\newtheorem*{remark}{Remark} 
\numberwithin{equation}{section}

\def\P{{P}}

\def\E{\mathbb{E}}
\def\d{\mathbf{d}}
\def\N{\mathbb{N}}

\def\Rp{\mathbb{R}_+}

\def\Z{\mathbf{Z}}
\def\z{\mathbf{z}}

\def\G{\mathcal{G}}
\def\C{\mathcal{C}}                 
\def\S{\mathcal{S}}

\def\Ep{\mathcal{E}}
\def\e{\boldsymbol e}
\def\ver{\boldsymbol v}

\def\dd{\ddot{d}}

\newcommand{\Tbd}[2]{T_{\sf{bd}}^{#1\mid #2}}
\newcommand{\Tn}[2]{T_{\sf{nei}}^{#1\mid #2}}
\newcommand{\Tbdx}[2]{\widetilde{T}_{\sf{bd}}^{#1\mid #2}}
\newcommand{\Tnx}[2]{\widetilde{T}_{\sf{nei}}^{#1\mid #2}}
\newcommand{\Tr}[2]{T^{#1\mid #2}}

\newcommand{\JT}[3]{J^{#1}_{(#2,#3^-)}}
\newcommand{\JE}[4]{\varepsilon^{#1}_{#2 (#3\rightarrow #4)}}

\def\deg{{\sf deg}}
\def\neig{{\sf nei}}

\def\deltanei{{\delta^{\textsf{nei}}}}

\def\parencite{\citep}
\def\textcite{\citet}

\begin {document}

\begin{frontmatter}
\title{On decomposable random graphs}
\runtitle{Decomposable random graphs}
\author{\fnms{Mohamad} \snm{Elmasri}\thanksref{a}\ead[label=e1]{mohamad.elmasri@mail.mcgill.ca}}
\runauthor{M. Elmasri}
\address{Department of Mathematics and Statistics \\ McGill University }

\thankstext{a}{ME was supported by the Fonds de recherche du Qu\'ebec - Nature et technologies (FRQNT) doctoral scholarship.}
\affiliation{McGill University}

\maketitle

\begin{abstract}
  Decomposable graphs are known for their tedious and complicated Markov update steps. Instead of modelling them directly, this work introduces a class of tree-dependent bipartite graphs that span the projective space of decomposable graphs. This is achieved through dimensionality expansion that causes the graph nodes to be conditionally independent given a latent tree. The Markov update steps are thus remarkably simplified. Structural modelling with tree-dependent bipartite graphs has additional benefits. For example, certain properties that are hardly attainable in the decomposable form are now easily accessible. Moreover, tree-dependent bipartite graphs can extract and model extra information related to sub-clustering dynamics, while currently known models for decomposable graphs do not. Properties of decomposable graphs are also transferable to the expanded dimension, such as the attractive likelihood factorization property. As a result of using the bipartite representation, tools developed for random graphs can be used. Hence, a framework for random tree-dependent bipartite graphs, thereupon for random decomposable graphs, is proposed.
\end{abstract}

\begin{keyword}
\kwd{decomposable graphs}
\kwd{junction trees}
\kwd{dimensionality expansion}
\kwd{bipartite graphs}
\kwd{Poisson processes}
\kwd{Markov chains}
\end{keyword}

\end{frontmatter}

\begin{center}
  \begin{minipage}{.85\linewidth}
    \setcounter{tocdepth}{2}
    \tableofcontents
  \end{minipage}
\end{center}

\section{Introduction}
The earliest introduction of decomposable graphs in statistics was by \cite{Lauritzen1980} and \cite{Laur83} as a generating class of decomposable log-linear models on multidimensional contingency tables. Decomposable graphs helped in reducing the number of factors without altering the maximum likelihood estimates, by collapsing factors belonging to the same connected components. Models using decomposable graphs have appeared since then in various topics in statistics. For example, the work of \cite{spiegelhalter1993} where decomposable graphs were used on Bayesian expert systems; \citet{cowell2006probabilistic} is a recent book on this topic. The work of \cite{Giudici01121999} and \cite{frydenberg1989decomposition} used the decomposability structure to factorize the likelihood for Bayesian model determination and mixed graphical interaction models, respectively. Overall, statistical applications of decomposable graphs can be categorized in two domains; in {\it graphical models}, as functional priors over covariance matrices, and in {\it Bayesian model determination}, as priors over hierarchies of model parameters. Few efforts in statistical literature exists beyond this scope.

The explicit interpretation of decomposability as conditionally dependent greatly simplifies the observational data likelihood. In particular, \cite{dawid1993} have showed that, if, and only if, a random variable \(X=(X_i)_{i<n}\) with a Markov distribution \(p\) and a conditional dependency abiding to a decomposable graph \(\G\), then the likelihood factorizes as
\begin{equation}\label{eq:factordist}
  p(X\mid \G) = \frac{\prod_{C\in\C}p(X_C)}{\prod_{S\in \S} p(X_S)},
\end{equation}
where \(\C\) and \(\S\) are graph subsets known as the maximal cliques and minimal separators, respectively, and \(X_A = \{X_i: i \in A \} \).

Despite the broad use of decomposable graphs in statistics, little work has been done on the sampling aspect. The lack of sampling methods is partly due to the complexity of testing for decomposability in large graphs, for example the size of the largest completely connected subgraph is still an open problem. And partly due to the lack of explicit methods that generate and quantify the combinatorial properties of decomposable graphs, such as, the space of junction trees or perfect orderings sequences. The recent notable work of \cite{thomas2009a} and \cite{stingo2015efficient} take a step in this direction, where both focus on updating, what is called, the junction tree of the graph, for faster mixing time. Nonetheless, computational complexity is still the largest obstacle. As noted by \cite{thomas2009a}, one of the best available tree search algorithms of a graph is by \cite{Tarjan:1984:SLA:1169.1179}, which is of the order \(\mathcal{O}(|\Theta| + |E|)\); where \(\Theta\) are the nodes and \(E\) are edges of the graph. Yet, for most dense graphs \(|E|\) is of the order \(\mathcal{O}(|\Theta|^2)\), and at best \(\mathcal{O}(|\Theta|)\) for sparse graphs. Following the terminology of \citet{bollobas2007metrics}, a graph of \(n\) nodes is called {\it dense} if the number of edges is of the order \(\mathcal{O}(n^2)\), and called {\it sparse} if its of the order \(o(n)\)

In parallel, the literature on {\it random graphs} have seen much development recently, where the interest is generally focused in modelling structural relational data in the form of a random \(d\)-arrays. Some known examples, the blockmodel \citep{wang1987stochastic}, latent distance model \citep{hoff2002latent}, infinite relational model \citep{kemp2006learning}, and many others. Refer to \citet{newman2003structure,newman2010networks} for a good introduction.

A general principle of random graph models is to assume an affinity parameter for each node in the network, governing its likelihood to form edges with other nodes. Affinities are hence seen as the drivers of the observational network structure, and modelling interest is mostly focused on their inference. Recent developments in random graphs points towards a unified modelling framework, based on the Aldous-Hoover and the Kallenberg representation theorems \citep{aldous1981representations,hoover1979relations,kallenberg1999multivariate}. Both representation theorems model random graphs as infinite objects, where a sample from such objects is through a random function indexed by node affinities. 

This work attempts to expand the use of decomposable graphs as general modelling objects. First, we address the complexity of updating decomposable graphs by treating them as projective objects from a special case of tree-dependent bipartite graphs. Moving in the space of decomposable graphs is done through simplified Markov update rules in the bipartite space, where graph nodes are conditionally independent given a latent tree. Hence, node updates can be done simultaneously. Like many other models for decomposable graphs, the proposed update step is iterative, updating the graph nodes given a latent tree and vice versa. Second, motivated by the Kallenberg representation theorem, the bipartite graph is modelled by means of a continuous point process driven by affinity parameters. Such representation allows for modelling of decomposable objects with random number of nodes. Properties that are hardly attainable in the decomposable graph form, such as the expected degree, come easy in the bipartite representation. Third, a likelihood factorization theorem, analogous to \eqref{eq:factordist}, with respect to the bipartite graph is given, which allow for direct application to graphical models.

The work is organized as follows, Section \ref{sec:preliminaries} introduces preliminaries on the combinatorial structure of decomposable graphs, the concept of perfect orderings, their relation to trees, and the Kallenberg representation theorem of random graphs. Section \ref{sec:GenerModel} introduces an alternative representation of decomposable graphs as projective objects from bipartite graphs. Permissible update moves in the bipartite space are illustrated. Section \ref{sec:decomposable-graphs-as-cont-process} introduces the model for decomposable random graphs by modelling the relevant bipartite graph as a continuous point process. Issues related to graph realizations are discussed, and a likelihood factorization theorem is given. Sections \ref{sec:sampling} and \ref{sec:sampling-T} illustrate an iterative sampling procedure for the proposed model. Section \ref{sec:expected-num-cliques-per-node} gives an exact expression of some expectation results. Section \ref{sec:examples} shows a some practical examples and few of their properties.

\section{Background}\label{sec:preliminaries}
\subsection{Decomposable graphs}\label{sec:IntroDecomGraph}
Let \(\G = (\Theta,E) \) be a simple undirected graph with a set of nodes \(\Theta = \{\theta_i\}_{i\in \N}\) and edges \(E = \{\{\theta_i, \theta_j\}\}_{i,j \in \N}\). A pair of nodes $\{\theta_i,\theta_j\} \in \Theta$ are adjacent if $\{\{\theta_i, \theta_j\}\}\in E$; the set notation is used since the edge \((\theta_i, \theta_j)\) is identical to \((\theta_j, \theta_i)\) in an undirected graph. However, this work uses the latter notation for simplicity. A graph \(\bar \G = (\bar \Theta, \bar E) \) is called a {\it subgraph} of \(\G\) if \(\bar \Theta \subset \Theta\) and (or) \(\bar E \subset E\). For simplicity, let \(\G(\bar \Theta)\) be the subgraph induced by \(\bar \Theta\subset \Theta\), where only edges between the nodes in \(\bar \Theta\) are included, similarly \(\G(\bar E)\) is a subgraph where only nodes forming edges in \(\bar E\) are included. A subset $C \in \Theta$ is said to be {\it complete} if every two distinct nodes in $C$ are adjacent, thus \(\G(C)\) is a complete subgraph of \(\G\) and is commonly called a {\it clique} of \(\G\). It is worth noting that subgraphs of cliques are also cliques, thus one can define a {\it maximal clique} to be a subgraph that cannot be extended by including any adjacent node while remaining complete. Consequently, all subgraphs of maximal cliques are also cliques, but not necessary maximal. 

The graph $\G$ is said to be {\it decomposable (chordal)} if, and only if, any cycle of four or more nodes has an edge that is not part of the cycle. Few equivalent definitions exist for decomposable graphs, one of which is the following.
\begin{mydef}\label{def:decom-graph}(Decomposable graphs, \citep{lauritzen1996}) A graph $\G=(\Theta, E)$ is decomposable if it could be partitioned into a triple $(A, S, B)$ of disjoint subsets of $\Theta$, such that
  \[ A \perp_\mathcal{G} B  \mid S, \quad \text{$S$ is complete}. \]
In other words, \(A\) is independent of \(B\) given \(S\).
\end{mydef}

A well known property of decomposable graphs is its {\it perfect ordering sequence (POS) of maximal cliques.} Denote the set of maximal cliques of $\G$ by $\C$, and let $|\C|=K$. Define a permutation $\pi:\{1,\dots, K\} \mapsto \{1, \dots, K\}$ such that,
\begin{equation} \label{eq:perfectorder}
\displaystyle
	H_{\pi(j)} = \bigcup_{i=1}^j C_{\pi(i)}, \quad S_{\pi(j)} = H_{\pi(j-1)} \bigcap C_{\pi(j)}, \quad \forall C_{i} \in \C.
\end{equation}

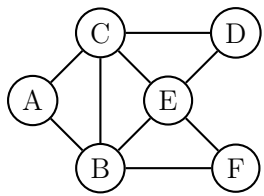
\begin{figure}[!ht]
	\centering
	\begin{tikzpicture}[scale=0.9, transform shape, thick]
        \tikzstyle{every node}=[font=\large]
        \node[circle,draw]  (ca) at(0,0){A};
	  	\node[circle,draw]  (cb) at(1,-1){B};
	  	\node[circle,draw]  (cc) at(1,1){C};
	  	\node[circle,draw]  (cd) at(3,1){D};
	  	\node[circle,draw]  (ce) at(2,0){E};
	  	\node[circle,draw]  (cf) at(3,-1){F};
	  	\draw (ca) -- (cb)--(cc)--(ce)--(cd)--(cc)--(ca);
	  	\draw(cb)--(ce)--(cf)--(cb);
	\end{tikzpicture}
	\caption{A decomposable graphs of 4 cliques of size 3; \{ABC, BCE, BEF, CDE\}.}
   \label{fig:FourCliques}
\end{figure}

Then, a sequence $\C_{\pi}= (C_{\pi(1)}, \dots,C_{\pi(K)})$ is called a POS of the maximal cliques of $\G$ if, and only if, for all $j>1$, there exist an $i < j$ such that $S_{\pi(j)} \subseteq C_{\pi(i)}$. The latter is known as the {\it running intersection property} (RIP) of the sequence. The set \(\S = \{S_{1}, \dots, S_{K}\}\) is called the {\it minimal separators} of $\G$, where each component \(\G(S_i)\) decomposes \(\G\) in a sense of Definition \ref{def:decom-graph}. While each maximal clique appears once in \(\C_{\pi}\), separators in \(\S\) could repeat multiple times, thus the naming of {\it minimal separators} as in the unique set of separators. The POS is a strong property, where a graph $\G$ is decomposable if, and only if, the maximal cliques of $\G$ could be numbered in a way that adheres to the RIP, thus forming a POS. Nonetheless, a decomposable graph could be characterized by multiple distinct POSs of its maximal cliques. For example, consider a graph formed of four triangles \{ABC, BCE, BEF, CDE\}, as shown in Figure \ref{fig:FourCliques}, with Table \ref{tab:FourCliquesPerfectOrder} listing three of its possible POSs.  
\begin{table}[!ht]

		\centering
		\caption{Possible prefect ordering of cliques of Figure \ref{fig:FourCliques}}
		\begin{tabular}{c c}
		perfect ordering & separators \\ $(C_{\pi(1)}, C_{\pi(2)}, C_{\pi(3)}, C_{\pi(4)})$ &$(S_{\pi(2)},S_{\pi(2)},S_{\pi(4)})$ \\
	 	\hline \\
		(ABC, BCE, CDE, BEF) & (BC, CE, BE) \\
		(CDE, BCE, BEF, ABC) & (CE, BE, BC) \\
		(BEF, BCE, ABC, CDE) & (BE, BC, CE) \\
		\end{tabular}
    \label{tab:FourCliquesPerfectOrder}
\end{table}

Despite the non-uniqueness of the POSs, \citet{lauritzen1996} has showed that the multiplicity of the minimal separators does not depend on the perfect ordering, implying a unique set of separators $\S$ across all POSs. Moreover, enumerating all POSs of a graph is directly related to enumerating the set of {\it junction trees}. A tree $T=(\C, \Ep)$ is called a junction tree of cliques of $\G$, or simply a junction tree, if the nodes of $T$ are the maximal cliques of $\G$, and each edge in \(\Ep\) corresponds to a minimal separator $S \in \S$. Here, with a slight abuse of notation, \(\C\) in \(T\) refers to a set of nodes indexing the maximal cliques of \(\G\). The junction tree concept is generally expressed in a broader sense, that is, for any collection \(\C\) of subsets of a finite set of nodes \(\Theta\), not necessary the maximal cliques, a tree \(T=(\C,\Ep )\) is called a junction tree if any pairwise intersection \(C_1 \cap C_2\) of pairs \(C_1, C_2, \in \C\) is contained in every node in the unique path in \(T\) between \(C_1\) and \(C_2\). Equivalently, for any node \(\theta \in \Theta\), the set of subsets in \(\C\) containing \(\theta\) induces a connected subtree of \(T\). The link between POSs and junction trees is more direct as shown by the following theorem.

\begin{theo}(Junction trees, \citep{cowell2006probabilistic}) A graph \(\G\) is decomposable if, and only if, there exists a junction tree of cliques.
\label{th:junction-tree}
\end{theo}

\begin{figure}[ht!]
  \centering
    \subfloat[a decomposable graph of a clique of size 3 ($C_2$), and two cliques of size 4 ($C_1,C_3$)]{
    \begin{tikzpicture}[scale=1, transform shape, thick]
      \node[circle,draw]  (ca) at(0,0){A};
      \node[circle,draw]  (cb) at(2,0){B};
      \node[circle,draw]  (cc) at(2,2){C};
      \node[circle,draw]  (cd) at(0,2){D};
      \node[circle,draw]  (ce) at(1,4){E};
      \node[circle,draw]  (cf) at(4,1){F};
      \draw (ca)-- (cb) -- (cd) -- (ca) -- (cc) -- (ce) -- (cd)--(cc)--(cb)--(cf);
      \draw (cd)--(cf)--(cc);
      \draw(0,1) --(0,1) node[anchor=east] {$C_1$};
      \draw(1,3) --(1,3) node[anchor=north] {$C_2$};
      \draw(4,1.5) --(4,1.5) node[anchor=south] {$C_3$};
        \end{tikzpicture}} \quad
  \subfloat[a connected bipartite graph between junction trees of cliques and perfect orderings]{
        \begin{tikzpicture}[scale=0.7, transform shape, thick]
          \node[circle,draw]  (c1) at(0,0){$C_1$};
          \node[circle,draw]  (c2) at(1.5,0){$C_3$};
          \node[circle,draw]  (c3) at(3,0){$C_2$};

          \node[circle,draw]  (c11) at(0,-1.5){$C_3$};
          \node[circle,draw]  (c21) at(1.5,-1.5){$C_1$};
          \node[circle,draw]  (c31) at(3,-1.5){$C_2$};
         \draw (c1)--(c2)--(c3); \draw (c11)--(c21)--(c31);
         \node  (o1)[right=2.5cm of c3]{$C_1, C_2,C_3$} ;
         \node  (o2)[below=0.5cm of o1]{$C_1, C_3,C_2$} ;
         \node  (o3)[below=0.5cm of o2]{$C_2, C_1,C_3$} ;
         \node  (o4)[below=0.5cm of o3]{$C_2, C_3,C_1$} ;
         \node  (o5)[below=0.5cm of o4]{$C_3, C_1,C_2$} ;
         \node  (o6)[below=0.5cm of o5]{$C_3, C_2,C_1$} ;

         \draw (3.5,0) -- (6,-1);\draw (3.5,0) -- (6,-4.5);\draw (3.5,0) -- (6,-5.5);\draw (3.5,0) -- (6,-3.4);
         \draw (3.5,-1.5) -- (6,-4.5);\draw (3.5,-1.5) -- (6,-1);\draw (3.5,-1.5) -- (6,0);\draw (3.5,-1.5) -- (6,-2.2);
         \draw node[left=0.5cm of c1]{$T_1$};
         \draw node[left=0.5cm of c11]{$T_2$};
         \draw (1.5,1) -- (1.5,1) node[anchor=south] {junction trees};
         \draw (7,1) -- (7,1) node[anchor=south]{perfect orderings};
         \end{tikzpicture}\label{fig:ex_bipartite-B}}
       \caption{A decomposable graph and its bipartite graph linking junction trees of cliques and perfect orderings.}\label{fig:ex_bipartite}
 \end{figure}
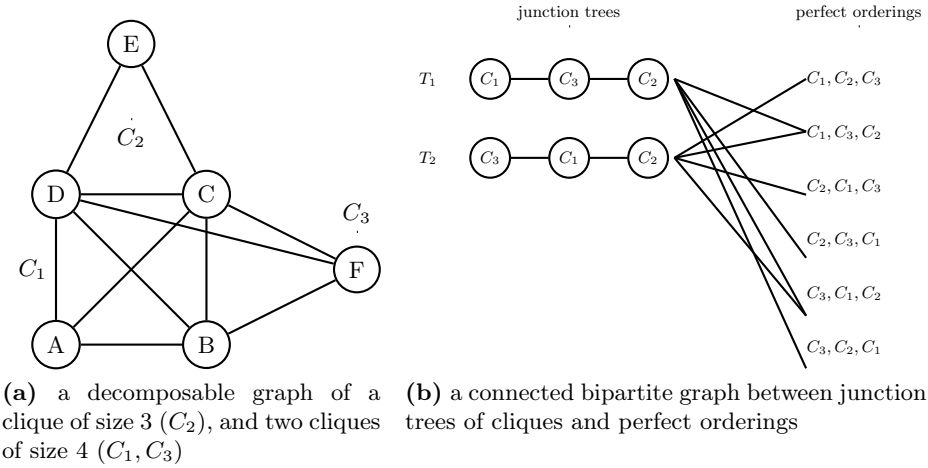

Despite the guaranteed existence of a junction tree, it is possible that a decomposable graph admits more than one unique junction tree, which is a direct consequence of the non-uniqueness of POSs. Nonetheless, since the set of separators is unique, the junction tree edge set \(\Ep\) is unique and characterizes all junction trees \citep{cowell2006probabilistic}. The connection between POSs and junction trees could be succinctly summarized in a bipartite network between both sets as shown in \citet{Hara06boundarycliques}, and illustrated by the example in Figure \ref{fig:ex_bipartite}. 

 The bipartite link in decomposable graphs between maximal cliques and junction trees play a central role in the generative process of the proposed model of this work. We use this dichotomy to move around the space of decomposable graphs by alternating between the two sets. Much of earlier work on decomposable graphs also focused on its junction tree representation, for its simplicity and computational efficiency.

 The next section briefly discusses the Aldous-Hoover and Kallenberg representation theorems \citep{aldous1981representations,hoover1979relations,kallenberg1999multivariate,kallenberg1990exchangeable}, and the representation of graphs as point processes.

 \subsection{Random graphs as point processes }\label{sec:kall-repr-rand}
A random matrix or a 2-array is a further generalization of a sequence of random variables. Much like infinite sequences, one can define an infinite matrix as
\begin{equation}
\xi = (\xi_{ij}) =\left ( \begin{array}{ccc}   
\xi_{11} & \xi_{12} & \dots \\    
\xi_{21} & \xi_{22} & \dots \\
 \vdots& \vdots & \ddots \end{array} \right ), 
\end{equation}
where the entries \((\xi_{ij})\) are random variables taking values, for example in \(\{0,1\}\). A random matrix is then called a {\it random graph} as it corresponds to an adjacency matrix of a graph. Hence, finite graph-valued data can be seen as partial observations from an infinite random matrix; much like the case of sequences of random variables. With a notion of exchangeability and some asymptotic results, one might be able to recover the generating distribution, up to some uncertainty, from the observed matrix; much like the law of large number acts on exchangeable sequences. The Aldous-Hoover representation theorem \citep{aldous1981representations, hoover1979relations} gives two notions of exchangeability associated with random matrices. When rows and columns represent the same set of objects, one might view exchangeability as a {\it joint} permutation of both rows and columns simultaneously.  If rows represent a different set of objects than columns, a {\it separate} permutation is then desirable.

Nonetheless, as discussed in \cite{orbanz2015bayesian} and based on the Aldous-Hoover representation theorem, random graphs represented by an exchangeable discrete 2-array are either trivially empty or dense. On the other hand, a notion of exchangeability based on continuous-space point processes can yield both, sparse and dense graphs, as shown by \citet{caron2014sparse,veitch2015class,borgs2014p}. The sparseness property of the model is crucial in many applications especially for real world large networks, as shown by \cite{newman2010networks}.

Representing a random graph as a point process on the continuous-space \(\Rp^{2}\) is achieved by embedding the graph nodes \((\theta_i)\) in the latter. Thus, the adjacency matrix \(\xi\) becomes a purely atomic measure on \(\Rp^2\) as 
\begin{equation}\label{eq:random-atomic-measures}
\xi =\sum_{i,j} z_{ij}\delta_{(\theta_i, \theta_j)},
\end{equation}
where \(z_{ij}=1\) if \((\theta_i, \theta_j)\) is an edge of the graph, otherwise \(z_{ij}=0\).

The focus of this work is on bipartite graphs, thus the following results of \citet{kallenberg1990exchangeable,kallenberg2006probabilistic} define the notion of separately exchangeable measures on \(\Rp^{2}\); let \(\Lambda\) denote the Lebesgue measure on \(\Rp\), \(\Lambda_D\) denote the Lebesgue measure on the diagonal of \(\Rp^2\).
\begin{theo}\label{th:Kallenberg-separately}(Kallenberg, separately exchangeable) A random measure \(\xi\) on \(\Rp^2\) is separately exchangeable if, and only if, almost surely

\begin{equation}\label{eq:Kallenberg-separately}
\begin{aligned}
  \xi =\quad &\sum_{i,j} f(\alpha, \vartheta_i, \vartheta'_j, U_{ij})\delta_{(\theta_i, \theta'_j)} \\
    + &\sum_{j,k} \Big ( g(\alpha, \vartheta_j, \chi_{jk})\delta_{(\theta_j,\sigma_{jk})} + 
                     g'(\alpha, \vartheta'_j, \chi'_{jk})\delta_{(\theta'_j,\sigma'_{jk})} \Big ) \\
+ &\sum_k \Big (l(\alpha, \eta_k)\delta_{(\rho_k, \rho'_k)}\Big ) \\
+ & \sum_j \Big ( h(\alpha, \vartheta_j)(\delta_{\theta_j} \otimes \Lambda) +  h'(\alpha, \vartheta'_j)(\Lambda \otimes \delta_{\theta'_j}) \Big ) + \gamma \Lambda^2.
\end{aligned}
\end{equation}
For some measurable function \(f\geq 0\) on \(\Rp^4\), \(g, g' \geq 0\) on \(\Rp^3\) and \(h, h', l, \geq 0 \) on \(\Rp^2\), some collection of independent uniformly distributed random variables \((U_{ij})\) on \([0,1]\), some independent unit rate Poisson processes \(\{(\theta_j, \vartheta_j)\}\), \(\{(\theta'_j, \vartheta'_j)\}\), \(\{(\sigma_{ij}, \chi_{ij})\}\) and \(\{(\sigma'_{ij}, \chi'_{ij})\}\) on \(\Rp^2\) and \(\{(\rho_j, \rho'_j, \eta_j)\}\) on \(\Rp^3\), for \(i,j \in N\), and some independent set of random variables \(\alpha, \gamma \geq 0\).  
\end{theo}

The primary part of the Kallenberg representation is the random function \(f\), which contributes most of the interesting structures of a graph; for more details refer to \citet[Sec. 4]{veitch2015class}. This work considers random atomic measures of the form in \eqref{eq:random-atomic-measures}, simplifying the expression in \eqref{eq:Kallenberg-separately} by considering all functions to be trivially zero, except \(f\). The next section introduces an alternative representation of decomposable graphs as projections from a special case of tree-dependent bipartite graphs. In the alternative representation the Markov update steps are greatly simplified.

\section{Decomposable graphs via tree-dependent bipartite graphs}\label{sec:GenerModel}

The building blocks of decomposable graphs are their maximal cliques. The smallest possible clique is a complete graph on two nodes (a stick). For our modelling purpose, we will regard the smallest possible clique to be an isolated node, where two isolated nodes form two maximal cliques, and connected they form a single maximal clique. Hence, an \(n\)-node graph can have a maximum of \(n\) maximal cliques, where all nodes are isolated, and a minimum of one single clique, where all nodes are connected forming an \(n\)-complete graph.

Relating the number of nodes to the range of possible cliques suggests that cliques can represent latent communities that are observed in the clique form by the attainment of node memberships. A decomposable graph is then an interaction between two sets of objects, the graph nodes and the latent communities. In the discrete case, out of the \(n\) possible communities of an \(n\)-node graph, only \(1\leq k\leq n\) communities are observed in the form of maximal cliques. The rest of \(n-k\) clique-communities are either latent with no visible node members or subgraphs of maximal cliques, either way they are unobserved. 

Let \(\G\) be a decomposable graph with \(T_{\G}\) being one of its junction trees. In classical settings, \(\G\) is modelled via its adjacency matrix and \(T_{\G}\) is a function of \(\G\), and research interest is in modelling the probability of node interactions.
\begin{description}
\item[Classical representation]
  \small
  \begin{equation}
    \textbf{Given:}\quad \G = (\Theta, E), \quad T_{\G} = f(\G), \quad
    \textbf{interest:}\quad \P( (\theta_{i}, \theta_{j}) \in E).
  \end{equation}
\end{description}

By separating the notion of nodes and maximal cliques, \(\G\) can be represented by a biadjacency matrix connecting the graph nodes to latent community nodes representing maximal cliques. Let \( \theta'_{1}, \theta'_{2}, \dots \in \Theta'\) be a set of nodes indexing the maximal cliques of \(\G\) and connected via the tree \(T = (\Theta', \Ep)\). Moreover, let \(\Z\) be the biadjacency matrix of \(\G\), where \(z_{ki}=1\) implies node \(\theta_{i}\) is a member of clique \(\theta'_{k}\), otherwise \(z_{ki}=0\). In essence, \(\Z\) represents a bipartite interaction between the two sets, \(\Theta'\) and \(\Theta\). Hence, the interest is in modelling the probability of node-clique interactions.
\begin{description}
\item[Alternative representation]
  \small
  \begin{equation}\label{eq:proposed-representation}
    \begin{aligned}
      \textbf{Given:}\quad \G = (\Theta, E), \quad T &= (\Theta', \Ep), \quad \Z =\big (\{\Theta', \Theta\},E_{Z} \big ),\\ \textbf{interest:}\quad &\P( z_{{ki}}=1).
    \end{aligned}
  \end{equation}
\(\G\) is a deterministic function of \(\Z\), since its adjacency matrix is
\begin{equation} \label{eq:mapping-adj-biadj}
  A = (a_{ij})_{ij}= \big ( \min\{\z^{\intercal}_{.i}\z_{.j},1\}\big)_{ij},
\end{equation}
where \(\z_{.i}\) is the \(i\)-th column of \(\Z\). Essentially, members of the same community, a row in \(\Z\), are connected in \(\G\).
\end{description}

\begin{remark}
  The notation \(\theta'_k\) is used interchangeably; for a subset of nodes of \(\Z\), for a subset of nodes of \(\G\) representing the maximal cliques, and for the nodes in \(T\). To avoid ambiguity, let the term "node(s)" refer to the graph nodes, and "clique-node(s)" to the nodes in \(T\), that is in \(\Theta'\). For simplicity, we will often use the term "clique \(\theta'_k\)" to refer to nodes of the maximal clique in \(\G\) represented by \(\theta'_{k}\), having the shorthand notation \(\G(\theta'_k)\). 
\end{remark}

The proposed representation assumes that an observed junction tree \(T_{\G}\) of \(\G\) is, in some way, a subtree of \(T\), since the maximal cliques \(\C\) of \(\G\) are a subset of \(\Theta'\). In fact, \(T_{\G}\) is a deterministic function of \(T\) given the set of maximal cliques. A more detailed description of the construction of \(T_{\G}\) from \(T\) is given later in Corollary \ref{coro:tg-from-t}.

\begin{figure}[ht!]
  \centering    
  \begin{tikzpicture}[->,>=stealth,thick]
    \node (g) at (0,0){\(\G\)};
    \node (z) at (5,0){\(\Z\)};

    \path[every node/.style={font=\tiny}]
    (g) edge[bend right] node [above] {injective in \(\Theta'\)} (z);

    \path[every node/.style={font=\tiny}]
    (z) edge[bend right] node [above] {surjective in \(\Theta'\)} (g);

    \path[every node/.style={font=\tiny}]
    (z) edge[right] node [above] {bijective in \(\Theta\)} (g)
    (g) edge node {} (z);
  \end{tikzpicture}
  \caption{Mapping relation between \(\G\) and \(\Z\).}
  \label{fig:domain-mapping}
\end{figure}
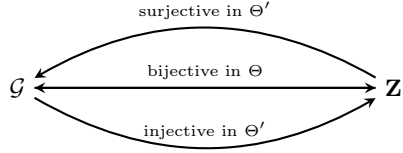

The link between \(\G\) and \(\Z\) in \eqref{eq:proposed-representation} is more intricate than the given simple expression of \eqref{eq:mapping-adj-biadj}. First, \(\Z\) is of a higher dimension than \(\G\), having an input of two sets of nodes. Moreover, the mapping from \(\Z\) to \(\G\) is bijective in the node domain and surjective in the clique domain, since the maximal cliques of \(\G\) are only a subset of \(\Theta'\). For example, it can be verified that for a single clique graph, \(\Z\) has one row in principle. Replicating this row, or adding rows representing clique subgraphs, would still map back to the same single clique graph by \eqref{eq:mapping-adj-biadj}. On the other hand, the inverse map from \(\G\) to \(\Z\) is bijective in the node domain, though injective in the clique domain. Figure \ref{fig:domain-mapping} summarizes the mapping relation between \(\G\) and \(\Z\).

The object \(\Z\) is then more general than a biadjacency matrix of a decomposable graph. In fact, one can define \(\Z\) independently from decomposable graphs, while retaining the relation in \eqref{eq:mapping-adj-biadj}. That is, a projection of \(\Z\) spans the space of decomposable graph. We term \(\Z\) as a tree-dependent bipartite graph and define it in Definition \ref{def:tree-dependent-graph}, while the following defines useful network functions used in this work.

\begin{mydef} \label{def:neig-deg}
  Denote \(\neig\) as the operator returning the set of neighbouring nodes, and \(\deg\) as the node degree. Such that, \(\neig(\theta_i, \G)\) are the neighbouring nodes of \(\theta_i\) in \(\G\) and \(\deg(\theta_i, \G) = |\neig(\theta_i, \G)|\) is the degree.  
\end{mydef}

\begin{mydef}[tree-dependent bipartite graph] \label{def:tree-dependent-graph}
  Let \(\Z = (\{\Theta',\Theta\}, E_{Z})\) be a bipartite graph connecting elements from the disjoint sets \(\Theta'\) and \(\Theta\). \(\Z\) is a tree-dependent bipartite graph if there exists a \(\Theta'\)-junction tree \(T=(\Theta', \Ep)\) of \(\Z\).  That is, for any pair \(\theta'_{1}, \theta'_{2} \in \Theta'\), \(\neig(\theta'_{1}, \Z) \cap \neig(\theta'_{2}, \Z) \subseteq \neig(\theta'_{k}, \Z)\), for every \(\theta'_{k}\) in the unique path in \(T\) between \(\theta'_{1}\) and \(\theta'_{2}\). 
\end{mydef}

Updating rules for \(\Z\) is less intuitive than what is classically proposed for \(\G\), though it comes with extra benefits. Classical models for decomposable graphs, such as the work of \cite{Green01032013}, adopt a tree-conditional iterative scheme for computational efficiency. Edges in \(\G\) are updated locally conditional on \(T_{\G}\), and iteratively \(T_{\G} \mid \G\) is updated using the relation in Figure \ref{fig:ex_bipartite}. A model for \(\Z\) ought to be iterative as well. The update rules for \(T\mid \Z\) are along the lines of the classical rules for \(\G\mid T\). Hence, the focus is in proposing a model for \(\Z \mid T\).

Sampling edges in a decomposable graph is highly dependent on the current configuration of the graph. Otherwise, (dis)connecting an arbitrary edge might hinder the graph undecomposable. This Markov dependency translates directly to \(\Z\). Given \(T\), sampling \(z_{ki}\) is highly dependent on the current configuration of \(\Z\), including that of \(z_{ki}\). \cite{Green01032013} have illustrated conditional (dis)connect moves on \(\G\mid T\) that ensure decomposability. Proposition \ref{prop:decom-graph-sub-maximal-treatment} below lists permissible moves in \(\Z \mid T\) that ensure \(\Z\) as a tree-dependent bipartite graph.

\begin{myprop}[Permissible moves in \(\Z\mid T\)] \label{prop:decom-graph-sub-maximal-treatment} Let \(\Z\) be a tree-dependent bipartite graph of Definition \ref{def:tree-dependent-graph}, with a \(\Theta'\)-junction tree \(T = (\Theta', \Ep)\). For an arbitrary node \(\theta_{i} \in \Theta\), let \(\Tr{}{i}\) be the subtree of \(T\) induced by the node \(\theta_i\) as
\begin{equation} \label{eq:induced-tree}
  \Tr{}{i} = T^{}\Big(\{\theta'_s \in \Theta': (\theta'_{s}, \theta_i) \in E_{Z}\} \Big ).\end{equation}

Moreover, let \(\Tbdx{}{i}\) be the boundary clique-nodes of \(\Tr{}{i}\), those of degree 1 (leaf nodes), and \(\Tnx{}{i}\) the neighbouring clique-nodes in \(T^{}\) to \(\Tr{}{i}\), as
\begin{equation}\label{eq:nei-bound-clique-basic-no-cond}
\begin{aligned}
& \Tbdx{}{i} = \Big \{\theta'_s \in \Theta': (\theta'_{s},\theta_{i})\in E_{Z}, \deg  (\theta'_s,\Tr{}{i})=1 \Big\},\\
& \Tnx{}{i} = \Big \{\theta'_s \in \Theta' : (\theta'_k, \theta'_s) \in \Ep, (\theta_{k}, \theta_{i})\in E_{z}, (\theta'_{s}, \theta_{i})\not\in E_{Z} \Big\}.
\end{aligned}
\end{equation}

Suppose \( \theta'_k \in \Tbdx{}{i}\bigcup \Tnx{}{i}\), let \(\Z'\) be the graph formed by one of the following moves:
\begin{equation}
\begin{aligned}
    &\textbf{connect: } &z_{ki}=1,\quad & \text{if}\quad \theta'_{k} \in \Tnx{}{i}, \\
    &\textbf{disconnect: } &z_{ki}=0,\quad & \text{if}\quad \theta'_{k} \in \Tbdx{}{i}.
  \end{aligned}
\end{equation}

Then \(\Z'\) is also a \(T\)-dependent bipartite graph.
\end{myprop}
\begin{proof} Note that the (dis)connect moves in Proposition \ref{prop:decom-graph-sub-maximal-treatment} do not alter \(T\). Thus, it is straightforward to verify that \(T\) is still the \(\Theta'\)-junction tree of \(\Z'\). Disconnect moves only affect leaf nodes of an induced tree. Hence, after a disconnection, pairwise intersection between any clique node with this leaf node is still contained in all clique nodes in their unique path in \(T\). Connect moves work similarly.
\end{proof}

\begin{mycoro}[\(T_{\G}\) a function of \(T\)]
A junction tree \(T_{\G}\) of a decomposable graph \(\G\) formed by \eqref{eq:mapping-adj-biadj} from a \(T\)-dependent \(\Z\) is a deterministic function of \(T\). It is the induced tree of \(T\), as \(T_{\G} =  T\big(\{\theta'_s \in \Theta': \theta'_{s} \text{ is maximal in } \G \} \big )\), if all nodes of maximal cliques are adjacent in \(T\). Otherwise, every non-maximal clique \(\theta'_{s}\) is contained in some maximal clique \(\theta'_{k}\) that is directly adjacent to it in \(T\), or through a path of other non-maximal cliques. Then, edges connected to \(\theta'_{s}\) in \(T\) can be rewired to \(\theta'_{k}\). This forms the tree \(T'\) where all maximal clique-nodes are adjacent. \(T_{\G}\) is then the induced tree of \(T'\).
\label{coro:tg-from-t}
\end{mycoro}

The boundary and neighbouring sets in \eqref{eq:nei-bound-clique-basic-no-cond} do not ensure that cliques remain maximal after a (dis)connect move. This is a direct consequence of the surjective mapping of the clique-node domain from \(\Z\) to \(\G\), as shown in Figure \ref{fig:domain-mapping}. Non-empty rows of \(\Z\) that represent sub-maximal cliques in \(\G\) would map to a maximal clique. To ensure that the bidirectional mapping between \(\Z\) and \(\G\) is injective, one must impose extra conditions on \(\Tbdx{}{i}\) and \(\Tnx{}{i}\) of Proposition \ref{prop:decom-graph-sub-maximal-treatment}. For \(\Tbdx{}{i}\), a boundary clique must stay maximal after a node's disconnection, and for \(\Tnx{}{i}\), a neighbouring clique must stay maximal after a node's connection. Formally, with abuse of notation, let \(\Tbd{}{i}\) and \(\Tn{}{i}\) be as in \eqref{eq:nei-bound-clique-basic-no-cond}, though with the extra imposed conditions as
\begin{equation}\label{eq:nei-bound-clique-basic}
\begin{aligned}
& \Tbd{}{i} = \Tbdx{}{i} \bigcap \Big \{\theta'_k \in \Theta': \theta'_k\setminus \{\theta_i\} \nsubseteq \theta'_s, s\neq k \Big\},\\
& \Tn{}{i} = \Tnx{}{i} \bigcap \Big \{\theta'_k\in \Theta' : \theta'_k\cup \{\theta_{i}\} \nsubseteq \theta'_{s}, s\neq k \Big\},
\end{aligned}
\end{equation}
where \(\theta'_k\setminus \{\theta_i\}\) refers to the subgraph formed by disconnecting the node \(\theta_{i}\) from clique \(\theta'_{k}\), and \(\theta'_k\cup \{\theta_i\}\) refers to the opposite, the subgraph formed by connecting node \(\theta_{i}\) to clique \(\theta'_{k}\). The extra imposed conditions in \eqref{eq:nei-bound-clique-basic} are clearly more restrictive than \eqref{eq:nei-bound-clique-basic-no-cond}, since a node's move now depends on neighbouring nodes. Regarding graphical models, the choice between the bidirectional-injective or the surjective-injective mapping is merely a modelling one. Section \ref{sec:likel-fact-with} gives a likelihood factorization theorem with respect to \(\Z\) that is equivalent to \eqref{eq:factordist}; disregarding the mapping choice.

The next section introduces a model for random decomposable graphs as realizations from continuous-time point processes in \(\Rp^{2}\). For issues related to the embedding in \(\Rp^{2}\), the bidirectional-injective mapping imposed by \eqref{eq:nei-bound-clique-basic} will be used. Nonetheless, a relaxation method of the imposed conditions in \eqref{eq:nei-bound-clique-basic} will be illustrated, though with a conceptually different viewpoint.

\section{Decomposable random graphs}\label{sec:decomposable-graphs-as-cont-process}

Drawing from the point process representation of graphs, introduced in Section \ref{sec:kall-repr-rand}, let \(\{(\theta_i, \vartheta_i)\}\) \(\{(\theta'_i, \vartheta'_i)\}\) be unit rate Poisson processes on \(\Rp^2\) representing the set of nodes \(\Theta\) and clique-nodes \(\Theta'\), respectively. Refer to \(\theta\) as the node location and \(\vartheta\) as the node weight. Given a tree \(T=(\Theta', \Ep)\), the tree-dependent bipartite graph \(\Z\), of Definition \ref{def:tree-dependent-graph}, now takes the form of a tree-dependent bipartite atomic measure on \(\Rp^2\), as
\begin{equation}\label{eq:biadj-measure-form}
\Z = \sum_{k,i} z_{ki}\delta_{(\theta'_k, \theta_i)}, 
\end{equation}
where \(z_{ki}=1\) if \((\theta'_{k}, \theta_{i})\) is an edge in \(\Z\), otherwise \(z_{ki}=0\). The decomposable graph \(\G\), which is now an atomic measure, is obtained from \(\Z\) as

\begin{equation} \label{eq:G-decomposable-graph}
\G = \sum_{i,j} \min\Big (\sum_k z_{ki}z_{kj}\delta_{(\theta'_k, \theta_i)}\delta_{(\theta'_k, \theta_j)},1 \Big )\delta_{(\theta_i, \theta_j)}.
\end{equation}

Let \(W:\Rp^2 \mapsto [0,1]\) be a simplified measurable function, analogous to \(f\) in \eqref{eq:Kallenberg-separately}. That is, for a uniform \([0,1]\) random array \((U_{ki})\), \(z_{ki}\) can be defined as \(z_{ki}:=\mathbb I \{U_{ki}\leq W(\vartheta'_k, \vartheta_i)\}\). Given the bidirectional-injective mapping, imposed by the neighbouring and boundary cliques in \eqref{eq:nei-bound-clique-basic}, we can accurately define the \(n+1\) Markov update step for \(z^{(n+1)}_{ki}\) conditional on the current configuration \(\Z^{(n)}\), as
\begin{equation}\label{eq:Update-step-full}
\P(z^{(n+1)}_{ki} = 1 \mid \Z^{(n)}, T) = 
\begin{cases}
 0 & \text{if } z_{ki}^{(n)}=0 \text{ and } \theta'_k \notin \Tn{(n)}{i}, \\
 1 & \text{if } z_{ki}^{(n)}=1 \text{ and } \theta'_k \notin \Tbd{(n)}{i},  \\
 W(\vartheta'_k, \vartheta_i) & \text{if } z_{ki}^{(n)}=1 \text{ and }\theta'_k \in \Tbd{(n)}{i}, \\ 
 W(\vartheta'_k, \vartheta_i) & \text{if } z_{ki}^{(n)}=0 \text{ and } \theta'_k \in \Tn{(n)}{i}.
\end{cases}
\end{equation}

Note that \(\theta'_k \in  \Tbd{(n)}{i}\) at step \(n\) only if \(\theta_i\) is a member of clique \(\theta'_k\), that is \(z^{(n)}_{ki}=1\). Similarly, \(\theta'_k\) is a neighbour to \(\Tr{(n)}{i}\) only if \(z^{(n)}_{ki}=0\). Otherwise, as in the first and second case in \eqref{eq:Update-step-full}, \(z^{(n+1)}_{ki}=z^{(n)}_{ki}\). This simplifies \eqref{eq:Update-step-full} to
\begin{equation}\label{eq:Update-step-simple}
\P(z^{(n+1)}_{ki} = 1 \mid \Z^{(n)}, T)= 
\begin{cases}
 W(\vartheta'_k, \vartheta_i) & \text{if }\theta'_k \in \Tbd{(n)}{i}\bigcup \Tn{(n)}{i}, \\ 
z^{(n)}_{ki} & \text{ otherwise.}
\end{cases}
\end{equation}

The form of \(W\) is still unspecified, and for it to be a sensible modelling object, the most general definition would require it to be at least measurable with respect to a probability space.

Under the Kallenberg representation theorem, a graph realization is seen as a cubic \([0,r]^2, r>0,\) truncation of \(\Rp^2\), and in that sense, the point process on \([0,r]^2\) might not be finite. In practice, a realization from a finite restriction is desired to be finite. \citet[Prop. 9.25]{kallenberg2006probabilistic} has given necessary and sufficient conditions for an exchangeable measure to be a.s. locally finite. Since we are not particularly focused on exchangeable random measures, yet still interested in the finiteness of a realization, the following definition simplifies Kallenberg's condition by focusing on a single-function atomic measure.

\begin{mydef}(locally finite) Let \(\xi\) be a random atomic measure on \(\Rp^2\), such that for a measurable random function \(W:\Rp^2 \mapsto [0,1]\), \(\xi\) takes the form
\begin{equation}\label{eq:atomic-measure-classic}
  \xi = \sum_{i,j} \mathbb I\{U_{ij} \leq W(\vartheta'_i, \vartheta_j)\} \delta_{(\theta'_i, \theta_j)},
\end{equation}
where \(\{(\theta'_i, \vartheta'_i)\}\) and \(\{(\theta_j, \vartheta_j)\}\) are two independent unit rate Poisson processes on \(\Rp^2\), and \((U_{ij})\) is a \([0,1]\) uniformly distributed 2-array random variables. The random measure \(\xi\) is a.s. locally finite if, and only if, the following conditions are satisfied:

\begin{enumerate}[(i)]
\item \(\Lambda\{\overline W_1=\infty\}=\Lambda\{\overline W_2 =\infty\}=0\),
\item \(\Lambda\{\overline W_1>1\}<\infty\) and \(\Lambda\{\overline W_2>1\}<\infty\),
\item \(\int_{\Rp^2} W(x,y)\mathbb I\{\overline W_1(y)\leq 1\}\mathbb I\{\overline W_2(x)\leq 1\} \d x\d y <\infty,\)
\end{enumerate}
where \(\overline W_1(y) = \int_{\Rp} W(x,y)\d x\) and  \(\overline W_2(x) = \int_{\Rp} W(x,y)\d y\), and \(\Lambda\) is the Lebesgue measure. In summary, if \(W\) is integrable then \(\xi\) is a.s. locally finite.
\label{def:locally-finite}
\end{mydef}

Thus far, we have introduced the general framework of the proposed model, the Markov update scheme at each step, and the required conditions on \(W\) to ensure a finite realization of the model. We will now give a formal definition of decomposable random graphs.

\begin{mydef}[Decomposable random graphs] A decomposable random graph \(\G\), projective random graph from a tree-dependent bipartite atomic random measure \(\Z\) taking the form in \eqref{eq:biadj-measure-form}, with the random function \(W: \Rp^2 \mapsto [0,1]\) satisfying the conditions in Definition \ref{def:locally-finite}, where \(\Z\) is constructed by means of a Markov process having the update steps of \eqref{eq:Update-step-simple}. A realization of such measure also takes the form of an \((r',r)\)-truncation as \(\Z_{r',r}=\Z(\, .\, \cap [0,r'] \times [0, r])\), for \(r', r > 0\).
\label{def:decomposable-graph-definition}
\end{mydef}

\begin{remark}
  The tree \(T=(\Theta', \Ep)\), connecting the clique-nodes \(\{(\theta'_{k}, \vartheta'_{k})\}\) in the \(T\)-dependent process \(\Z\), can be seen as the limit of junction trees of realizations from \(\G\) as more nodes enter the truncation.
\end{remark}

The sampling notion of an \((r',r)\)-truncation mentioned in Definition \ref{def:decomposable-graph-definition} is not yet fully discussed, in particular, how it assures decomposability with scaling of \(r\) or \(r'\). The next section formalizes this notion, where certain issues relating to the bidirectional mapping choice between \(\Z\) and \(\G\) are illustrated; with some proposed solutions. The following definition introduce useful graph functions and notations used in this work.

\begin{mydef} Denote the operators \(\ver\) and \(\e\) as node and edge sets of graph like structures, respectively. Such that, \(\ver(\G(x)) \in \Theta\) are the nodes of the subgraph \(\G(x)\), and \(\e(\G(x)) \in E\) are the edges. Similar notations for \(\Z\) and \(T\). To distinguishing between nodes and clique-nodes in \(\Z\), denote \(\ver_n(\Z(y)) := \ver(\Z(y))\setminus\Theta'\) as the set of graph nodes, and \(\ver_c(\Z(y)) := \ver(\Z(y))\setminus \Theta\) as the set of clique-nodes.
\label{def:node-set-and-edge-set}
\end{mydef}

\subsection{Finite graphs forming from domain restriction}\label{sec:finite-graph-restrictions}
Under the Kallenberg representation theorem, a realization from \(\Z\) is seen as the cubic restriction \([0,r']\times [0,r]\) of \(\Rp^2\), where clique-nodes and nodes within the cube are visible in the graph form only if they appear in some edge in \(\Z_{r',r}=\Z(\, .\, \cap [0,r'] \times [0, r])\). In this case, refer to the appearing clique-nodes and nodes as {\it active}.

There is a clear ambiguity relating to the influence of each domain restriction on the other, specially due the Markov formation of the graph. Nonetheless, if we neglect for a moment the formation method, and regard \(\Z\) as infinite fixed object sampled by the \((r',r)\)-truncation as \(\Z_{r',r}\), there is still doubt on how an achieved realization forms a decomposable graph. For example, a random embedding of clique-node locations \(\theta'_1, \theta'_2, \dots\) in \(\Rp\) can result in an empty realization even for large values of \(r'\); influenced by the inter-dependency between clique-nodes in \(T\). Moreover, there is still no promise that active clique-nodes, if any, form maximal cliques, since the truncation can cut through the clique formation, hindering them non-maximal. Essentially, what is required is that a realization from \(\Z\) represent a decomposable graph with a junction tree as subtree of \(T\); not necessary completely connected.

A simple way to address those issues is to ensure that the restriction point \(r\) is magnitudes larger than \(r'\), to allow enough active nodes such that all active clique-nodes are maximal. Essentially, scale the truncation size while ensuring that the following {\bf A0} set is empty.
{\footnotesize
\begin{equation}\label{eq:epmty-set-req}
  {\bf A0}:=
\Big \{\theta'_k<r': \Z_{r',r}(\{\theta'_k\}\cap \, .\,) \subseteq \Z_{r',r}(\{\theta'_s\}\cap\, .\,),\, \theta'_s<r', s \neq k, \theta'_k \text{ is active} \Big \}
\end{equation}}

Note that \(\theta'_s\) need not be active in \eqref{eq:epmty-set-req}, as for non-active cliques \( \Z_{r',r}(\{\theta'_k\}\cap\, .\,)= \{\emptyset\}\). The conditions in {\bf A0} are the same conditions added to the boundary and neighbouring sets in (\ref{eq:nei-bound-clique-basic}).

\begin{remark}
Allowing \(r\) to be much larger than \(r'\) relates directly to the notion discussed at the beginning of Section \ref{sec:GenerModel}, that the set of maximal cliques of \(T\) are only partially observable given the nodes. 
\end{remark}

\begin{figure}[htp!]
  \centering    
\subfloat[\(T\)-dependent process \(\Z\)]{\includegraphics[width=0.56\textwidth]{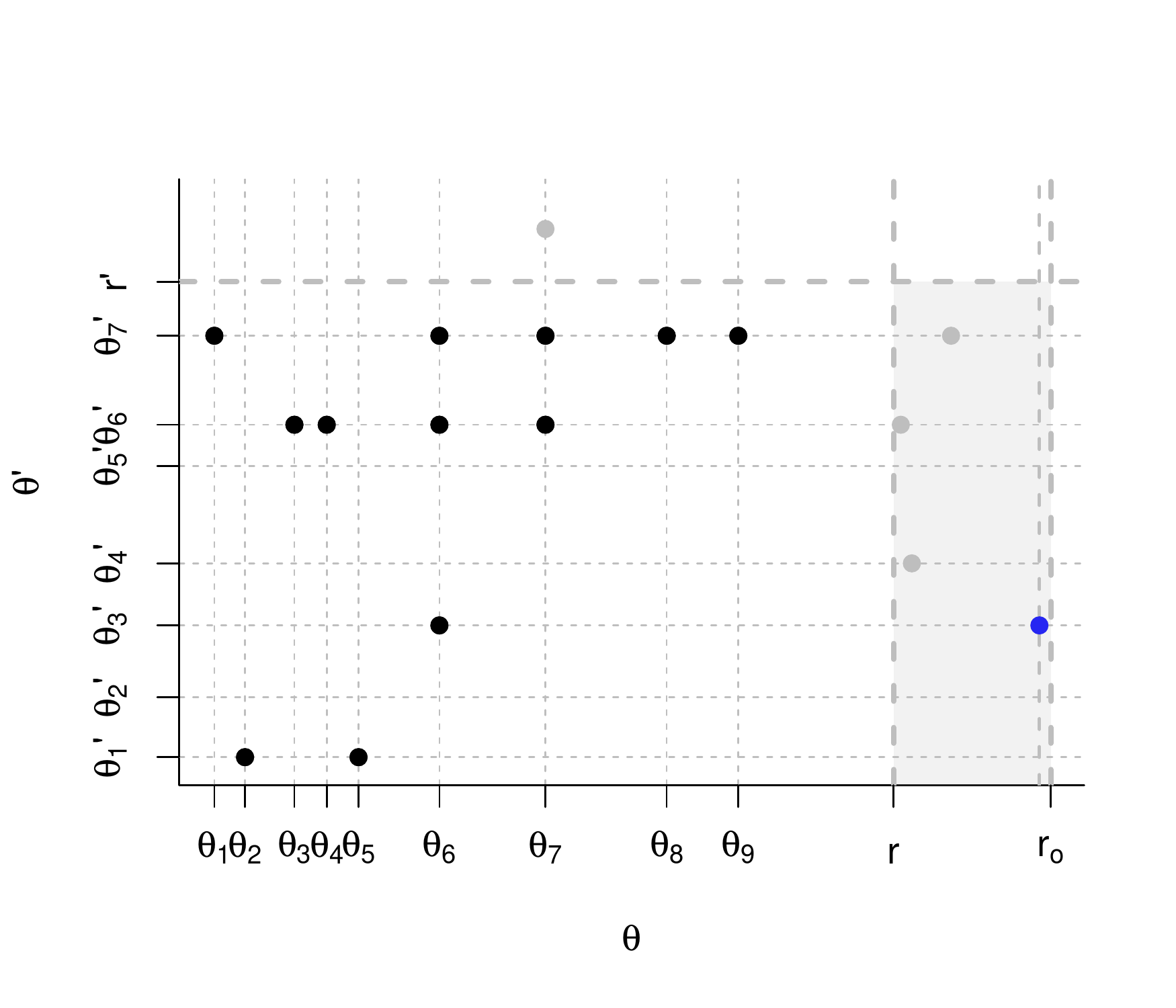}\label{fig:sub-A-decom-example1}} \hspace{-5.5em}
\subfloat[latent tree \(T\)]{\includegraphics[width=0.56\textwidth]{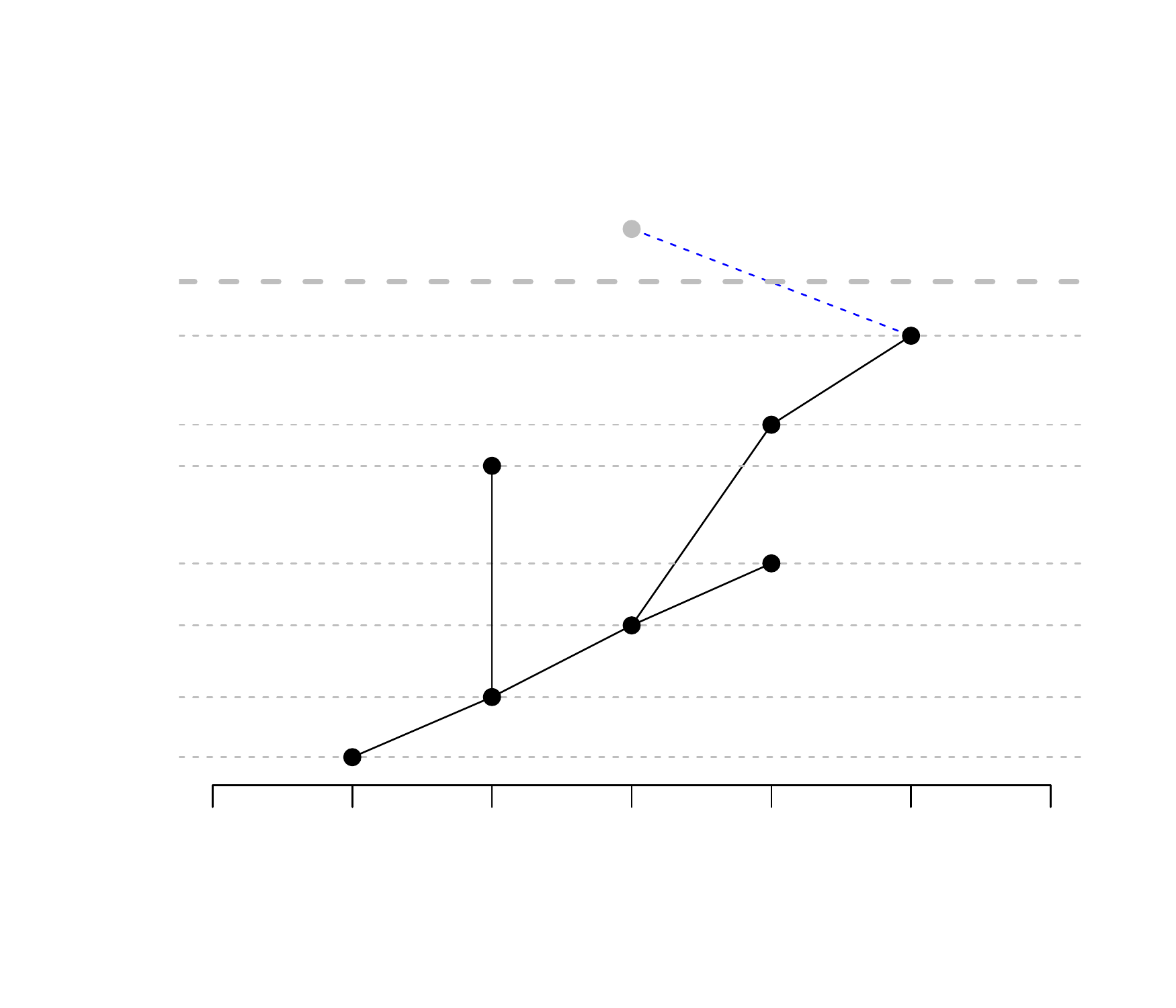}\label{fig:sub-B-decom-example1}}
\vspace{2em}

\subfloat[realization with an edge-greedy node]{\includegraphics[width=0.50\textwidth]{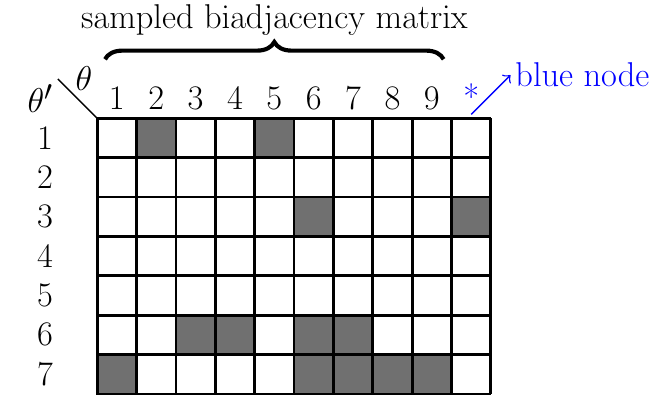}\label{fig:sub-C-decom-example1}}
\quad \quad
	\subfloat[mapped decomposable graph]{
      \begin{tikzpicture}[scale=0.7, transform shape, thick]
	  \tikzstyle{every node}=[font=\large]
	  	\node[circle,draw]  (ca) at(-1,2){2};
	  	\node[circle,draw]  (cb) at(-1,0){5};
       
	  	\node[circle,draw]  (cc) at(1,0){3};
	  	\node[circle,draw]  (cd) at(3,0){4};
	  	\node[circle,draw]  (ce) at(3,2){6};
	  	\node[circle,draw]  (cf) at(1,2){7};
        
	  	\node[circle,draw]  (cg) at(0.5,3.5){1};
	  	\node[circle,draw]  (ch) at(2,4.5){8};
	  	\node[circle,draw]  (cj) at(3.5,3.5){9};

        \node[circle,draw, blue]  (ck) at(5,2){*};
        \draw[blue] (ce) -- (ck);
        \draw (ca) -- (cb);
        \draw(cc) -- (cd) -- (ce) -- (cf) --(cc)-- (ce);
        \draw (cd) --(cf) -- (cg) --(ch) --( cj) --(ce) --(cg)--(cj)--(cf)--(ch) --(ce);
	\end{tikzpicture}\label{fig:sub-D-decom-example1}}
\caption{A realization of a decomposable graph in \ref{fig:sub-D-decom-example1} from the point process in \ref{fig:sub-A-decom-example1} and the tree \ref{fig:sub-B-decom-example1}. The grey area in \ref{fig:sub-A-decom-example1} is the edge-greedy partition $(r,r_o]$, where only one extra node (in blue) was needed to guarantee all active cliques are maximal, since $\theta'_3$ is a subset of $\theta'_6$ and $\theta'_7$, as shown in \ref{fig:sub-C-decom-example1}, which is the biadjacency matrix of active (clique-)nodes.}
\label{fig:decom-example1}
\end{figure}
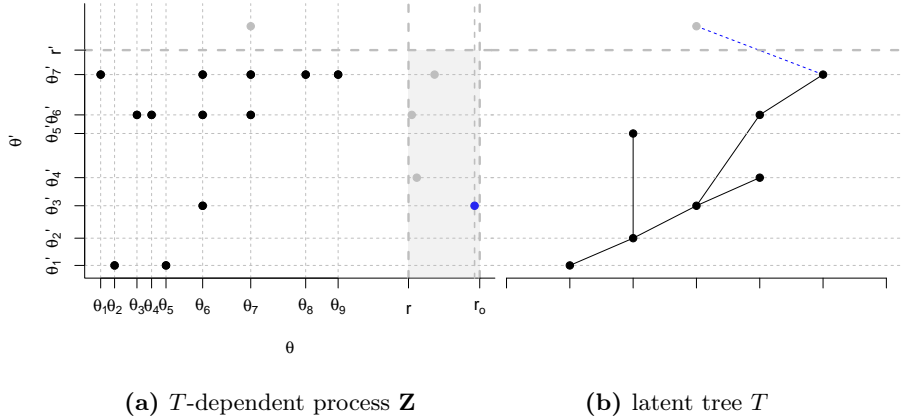
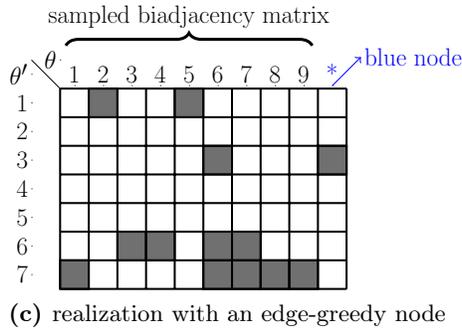
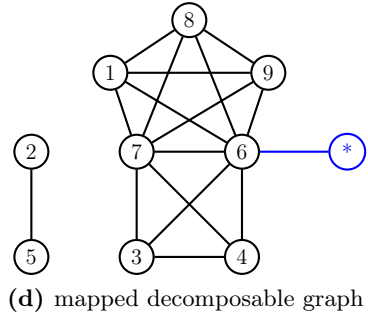

To scale freely the \((r',r)\)-truncation, while ensuring {\bf A0} is empty, one can, extend the truncated \(\theta\)-domain by an {\it "edge-greedy"} partition \((r, r_o]\) as \([0,r]\cup (r, r_o]\). Rather than trimming all external edges connecting from \((r, r_o]\) to \([0,r']\), as done with edges outside the cube \([0,r']\times[0,r]\), allow a maximum of a single edge per node \(\theta_i \in (r,r_o]\) to connect to any active clique-node \(\theta'_k \in [0,r']\). This connection is allowed only if it causes \(\Z_{r',r_o}(\{\theta'_k\} \cap \, .\,)\) to be maximal, while it is not in \(\Z_{r',r}(\{\theta'_k\} \cap \, .\,)\). Figure \ref{fig:decom-example1} illustrates this process on a \([0,r'] \times [0, r]\) cube with an edge-greedy partition \((r, r_0]\). The realized decomposable graph is also shown, where one extra node, \(\theta^*\), is included to clique \(\theta'_3\) to insure the set in \eqref{eq:epmty-set-req} is empty.

Condition {\bf A0} is not such a computational burden. Arriving at where to allow an edge in the edge-greedy partition can be done through matrix operations, by comparing the off-diagonal to the diagonal entries of \(\Z_{r',r} \Z_{r',r}^\top\). This comparison is only needed against the neighbouring clique-nodes, thus of linear complexity with respect to the number of active clique-nodes. Nonetheless, a simpler solution is possible by an identity matrix augmentation, as shown in the next subsection

\subsubsection{Augmentation by an identity matrix}

\begin{figure}[htp!]
  \centering    
\subfloat[][augmented bipartite matrix.]{\includegraphics[width=0.45\textwidth]{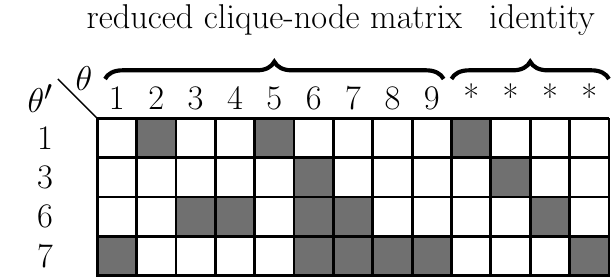}\label{fig:decom-example1-identity-A}}\quad\quad
\subfloat[mapped decomposable graph]{
  \begin{tikzpicture}[scale=0.7, transform shape, thick]
        \tikzstyle{every node}=[font=\large]
	  	\node[circle,draw]  (ca) at(-1,2){2};
	  	\node[circle,draw]  (cb) at(-1,0){5};
        \node[circle,draw]  (css) at(0,1){*};
	  	
        \node[circle,draw]  (cc) at(1,0){3};
	  	\node[circle,draw]  (cd) at(3,0){4};
	  	\node[circle,draw]  (ce) at(3,2){6};
	  	\node[circle,draw]  (cf) at(1,2){7};
        \node[circle,draw]  (cs) at(4,1){*};
        
	  	\node[circle,draw]  (cg) at(0,4){1};
	  	\node[circle,draw]  (ch) at(2,5.5){8};
	  	\node[circle,draw]  (cj) at(4,4){9};
	  	\node[circle,draw]  (csss) at(2,3.5){*};

        \node[circle,draw]  (ck) at(5,2){*};
        \draw (ce) -- (ck);
        \draw (ca) -- (cb) -- (css) -- (ca);
        \draw(cc) -- (cd) -- (ce) -- (cf) --(cc)-- (ce);
        \draw (cd) --(cf) -- (cg) --(ch) --( cj) --(ce) --(cg)--(cj)--(cf)--(ch) --(ce);

        \draw (cc) -- (cs) -- (cd);
        \draw (cf) -- (cs) -- (ce);
        \draw (ce) -- (csss) -- (cf); 
        \draw (cg) -- (csss) -- (ch) ;
        \draw (cj) --(csss);
	\end{tikzpicture}\label{fig:decom-example1-identity-A}}
\caption{Relaxation of \eqref{eq:epmty-set-req} by removing the empty rows in the realization of Figure \ref{fig:sub-C-decom-example1} and augmenting the results with an identity matrix.}
\label{fig:decom-example1-identity}
\end{figure}
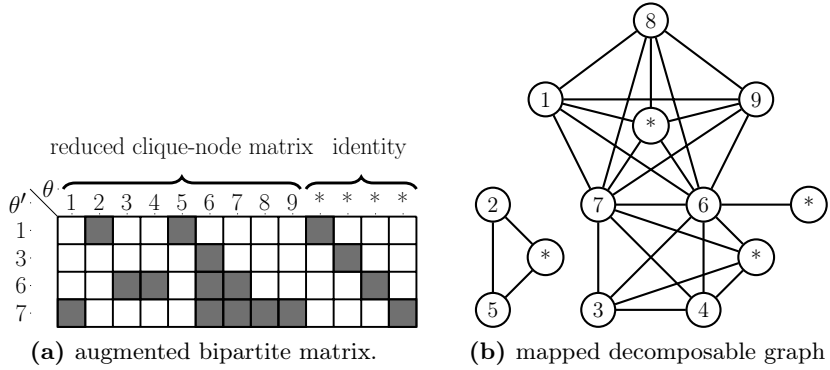

Clearly, rather than checking {\bf A0}, one can one simply augment a realization by an identity matrix, after the removal of empty rows. This, post-sampling procedure, artificially adds a maximum \(N_c\) extra nodes to the graph, each node is connected to a single unique clique, thus uniquely index the clique set. Figure \ref{fig:decom-example1-identity} summarizes this process for the realization in Figure \ref{fig:sub-C-decom-example1}. Such augmentation is a natural consequence of the used framework and the edge-greedy partition. In a sense, given the \((r',r)\)-truncation method, for any realization over \(\Z_{r',r}\), with probability 1, there exists an \(r_o>r\) such that the edge-greedy partition \((r,r_o]\) embeds an identity matrix. To show this, we will extend the results of \citet{veitch2015class} concerning the degree distribution of the Kallenberg exchangeable graphs to the case of a bipartite measure.

Consider the random bipartite atomic measure \(\xi\) of Definition \ref{def:locally-finite}. A realization from \(\xi\) is also an \((r',r)\)-truncation as \(\xi_{r',r} = \xi(\, .\, \cap [0,r']\times [0,r])\) where only edge-connected nodes are visible. The construction of \(\xi\) differs from that of the decomposable graph in Definition \ref{def:decomposable-graph-definition}, where the latter is conditioned on a latent tree structure while the former is not. Given \(\xi_{r',r}\), one can define the degree distribution of any point in the domain of the \(r\)-truncated Poisson process \(\Pi_r\), or any of the two domains by symmetry. Let
\begin{equation}\label{eq:degree-function}
\deg(\theta, \xi_{r', r}):=\deg((\theta, \vartheta), \Pi_r, \Pi'_{r'}, (U_{ij})),
\end{equation}
be the degree of the point \((\theta, \vartheta)\) in the domain of \(\Pi_r\), with the left-hand side being a simplified notation. Nonetheless, the probability that \((\theta, \vartheta) \in \Pi_r\) is 0, thus as noted by \citet{veitch2015class} and discussed more generally in \citet{chiu2013stochastic}, this conditioning is ambiguous and ill formulated. Nonetheless, a version of the required conditioning can be obtained by the Palm theory of measures on point sequences. The Slivnyak-Mecke theorem states that the distribution of a Poisson process \(\Pi\) conditioned on a point \(\{x\}\) is equal to the distribution of \(\Pi \cup \{x\}\); with this we characterize the degree distribution in the following lemma.

\begin{mylemma} For a bipartite measure \(\xi_{r',r}\), defined as in \ref{def:locally-finite}, with a non-random \(W\), the degree distribution of a point \((\theta, \vartheta) \in \Rp^2\), \(\theta<r\), is \(\deg(\theta,\Pi_r \cup \{(\theta, \vartheta)\}, \xi_{r',r}) \stackrel{\text{d}}{\sim}\text{Poisson}(r'\overline W_1(\vartheta)), \) and by symmetry of construction, \(\deg(\theta', \Pi'_{r'} \cup \{(\theta', \vartheta')\},\xi_{r',r}) \stackrel{\text{d}}{\sim}\text{Poisson}(r\overline W_2(\vartheta')) \).
\label{lem:degree-dist-regular-biadjacency} 
\end{mylemma}

\begin{proof} Since \((\theta, \vartheta) \in \Pi_r\) with probability 0, for a fixed \(W\), using the Palm theory we have
\begin{equation} \label{eq:degree-dist-bipartite}
\deg(\theta, \Pi_r \cup \{(\theta, \vartheta)\},\xi_{r',r}) = \sum_{(\theta'_k, \vartheta'_k) \in \Pi'_{r'}} \mathbb I\{ U_{\theta'_k, \theta} \leq W (\vartheta'_k, \vartheta)\}.
\end{equation}

By Definition \ref{def:locally-finite}, \(W\) is a.s. finite, thus by a version of Campbell's theorem \parencite[ch 5.3]{kingman1993poisson}, the characteristic function of \eqref{eq:degree-dist-bipartite} is 
\begin{equation}
\begin{aligned}
\E\Big [\exp\Big (it \deg \big (\theta,\; .\;\big)\Big ) \Big] 
&=\exp \Big (\int_{\Rp}\int_{[0,1]}(1 - e^{it\mathbb I\{ u \leq W (x, \vartheta)\} })r'\d u\d x \Big) \\
&= \exp\big(r' \overline W_1(\vartheta)(e^{it}-1) \big), 
\end{aligned}
\end{equation}
where \(\overline W_1(y) = \int_{\Rp} W(x,y)\d x\). The result follows similarly for the second domain. For a random \(W\), the same result can be achieved by conditioning.
\end{proof}

With a well-defined degree distribution for \(\xi_{r'.r}\), we can show that the identity matrix augmentation, in Figure \ref{fig:decom-example1-identity}, is a natural consequence.

\begin{myprop} \label{prop:identity-always-exists} Following Definition \ref{def:decomposable-graph-definition}, for a realization from \(\Z_{r',r}\), let \(\Theta'_{r'}\) be the set of active clique-nodes with \(|\Theta'_{r'}|>1\). Then, with probability 1, there exists an \(r_o>r\), such that each \(\theta'_k \in \Theta'_{r'}\) is indexed by a unique node \(r<\theta_{\pi(k)}<r_o\). Hence, the partition \((r,r_o]\) embeds an identity matrix. 
\end{myprop}

\begin{proof}
Given realization from \(\Z_{r,'r}\), index the almost surely finite set of active clique-nodes as \(\theta'_1, \theta'_2, \dots, \in \Theta'_{r'}\). For \(t>0\), let
\[ \begin{aligned} Y_t^{(k)} &:= |\e(\Z_{r',r+t}(\{\theta'_k\}\cap \, .\, ))| - |\e(\Z_{r',r}(\{\theta'_k\}\cap \, .\, ))| \\
&:= \deg(\theta'_k, \Z_{r', r+t}) - \deg(\theta'_k,\Z_{r',r}),
\end{aligned}
\]
as the degree of the \(k\)-th active clique \(\theta'_k \in \Theta'_{r'}\) over the partition \((r, r+t]\). For a finite \(t\), \(Y_t^{(k)}\) is an almost surely non-negative finite process, by finiteness of the generating measure (Definition \ref{def:locally-finite}). For the filtration \(\mathcal F := \sigma (\{(\theta'_k, \vartheta'_k)\}, T) \), define \(\tau^{(k)}\) to be the stopping time of the event that an edge appears between a node and \(\theta'_{k}\), in a unit interval, while no edge in the same interval appears for the rest of the active clique-nodes. Formally,
{\small
\begin{equation}
  \tau^{(k)} := \min\Big \{t\in \N : \{Y_{t + 1}^{(k)} - Y_t^{(k)}>0\}\bigcap_{s\neq k}\{Y_{t+1}^{(s)}-Y_t^{(s)} =0\}, \theta'_s \in \Theta'_{r'}\Big\}.
\end{equation}
}
It suffices to show that \(\tau^{(k)}< \infty \) with probability 1 for each \(\theta'_k \in \Theta'_{r'}\) and then take \(r_o = \max_k(\tau^{(k)})\). By conditioning on \(T\), \((Y_t^{(k)})_{k}\) are not independent and do not yield an accessible distribution. Nonetheless, if we let \((\widetilde Y_t^{(k)})_{k}\) represent the analogous process though under \(\xi_{r',r}\) of Definition \ref{def:locally-finite}, then \((\widetilde Y_t^{(k)})_{k}\) are independent with a well-defined distribution (Lemma \ref{lem:degree-dist-regular-biadjacency}). Moreover, for each \(k\), \(Y_t^{(k)}\) is dominated by \(\widetilde Y_t^{(k)}\) as \(Y_t^{(k)}\leq \widetilde Y_t^{(k)}\), since the latter could be seen as induced by an infinite complete graph \(K_\G\), where \(T\subset K_\G\). For an analogous filtration \(\widetilde{\mathcal F} := \sigma (\{(\theta'_k, \vartheta'_k)\}, K_\G) \), and the stopping time \(\widetilde \tau^{(k)}\) under \((\widetilde Y_t^{(k)})_k\) we have 
{\small
 \begin{equation}
\begin{aligned}
\P(\tau^{(k)} \geq n) &\leq \P(\widetilde \tau^{(k)} \geq n) \leq \frac{1}{n}\E [\widetilde \tau^{(k)} \mid \widetilde{ \mathcal{F}}]\\ 
&=\frac{1}{n}\E \Big [\sum_{t\geq 1} t \mathbb I\{ s<t: s\neq \widetilde \tau^{(k)}\} \mathbb I \{\widetilde \tau^{(k)}=t \}  \mid \widetilde {\mathcal{F}}\Big ]\\
&\leq \frac{1}{n} \sum_{t\geq 1} t\Big [\prod_{i=1}^{t-1}1-\P\Big (\widetilde Y_{i + 1}^{(k)} - \widetilde Y_i^{(k)}>0\Big)\P\Big(\bigcap_{s\neq k}\{\widetilde Y_{i+1}^{(s)}-\widetilde Y_i^{(s)} =0\}\Big ) \Bigg]\\
&\leq \frac{1}{n} \sum_{t\geq 1} t\Bigg[ 1- \exp{\Big (-\sum_{s\neq k} \overline W_2(\vartheta'_s)\Big )}\Bigg]^{t-1} \\
&= \frac{1}{n} \exp\Big(2\sum_{s\neq k}\overline W_2(\vartheta'_s)\Big) \longrightarrow 0 \quad \text{as} \quad n \longrightarrow \infty .
\end{aligned}
\end{equation}
}
The inequalities above are a result of the Markov inequality, the independence of \((\widetilde Y_t^{(k)})_k\), the removal of the first probability in the third line, the direct application of the geometric series sum, and finally by condition (i) in Definition \ref{def:locally-finite}. The proof could be also achieved by the Borel-Cantelli Lemma.
\end{proof}

This section discussed theoretical implications resulting from the \((r',r)\)-truncation method. It addressed the issue of how a random truncation of the node domain can cause cliques to be non-maximal, and what conditions are needed to ensure that all active cliques are maximal. To ease the computational complexity of the needed conditions, a selective post-sampling procedure was proposed. The edge-greedy partition adds a dummy node to every non-maximal active clique. With probability 1, the point process representation guarantees that there exists an extended partition embedding an identity matrix, a unique node for each active clique. This justifies theoretically the proposed post-sampling procedure.

The next section links the results obtained for the \(T\)-dependent bipartite process \(\Z\) to the likelihood factorization property of \(\G\).

\subsection{Likelihood factorization with T-dependent processes}\label{sec:likel-fact-with}
Most statistical applications of decomposable graphs are in the field of graphical models, where the interest is in inferring the Markov dependencies among variates. In this context, the structural dependency of a random variable is assumed to follow \(\G\). Section \ref{sec:finite-graph-restrictions} addresses issues arsing from representing \(\Z\) as a point process, how non-maximal cliques can occur by randomly truncating the node domain, and a post-sampling correction procedure. Moreover, in \eqref{eq:Update-step-simple}, the restrictive boundary and neighbouring sets of \eqref{eq:nei-bound-clique-basic} were used to ensure the bidirectional-injective mapping between \(\Z\) and \(\G\), rather than the simpler sets in \eqref{eq:nei-bound-clique-basic-no-cond}. In both cases, the objective was a way to handle non-maximal active cliques. Regarding graphical models, do non-maximal active cliques in \(\Z\) alter the shape, hence the likelihood, of \(\G\)? In other words, is the post-sampling procedure necessary? Moreover, is there a likelihood factorization with respect to \(\Z\), analogous to that of \eqref{eq:factordist}, allowing direct modelling with \(\Z\)? 

\begin{theo}[Likelihood factorization with respect to \(\Z\)] \label{th:Z-likeli-factor}
  Let $\Z$, \(T\), and \(\G\) be as in Definition \ref{def:decomposable-graph-definition}, and let $X = (X_i)_{i<N_{v}}$ be a random variable with a Markov distribution $p$ and conditional dependency abiding to $\G$, then the likelihood of $X \mid \Z$ is
\begin{equation}\label{eq:Z-likelihood-factorization}
p(X \mid \Z)= \frac{\prod_{\theta'_k \in \Theta'} p(X_{\theta_k})}{\prod_{(\theta'_k, \theta'_j)\in \Ep} p(X_{\theta'_k\cap \theta'_j})}.
\end{equation}
In fact, \(p(X \mid \Z)  = p(X \mid \G)\) of \eqref{eq:factordist}. Moreover, the above holds disregarding the mapping choice.
\end{theo}

\begin{proof}
  Assuming that not all $\theta'_k \in \Theta'$ are maximal, we will show that every non-maximal clique in the numerator of \eqref{eq:Z-likelihood-factorization} cancels out with an equivalent factor in the denominator, leaving the minimal separator set $\mathcal S$ of \(\G\) as in (\ref{eq:factordist}). First, note that $X_\emptyset=\emptyset$, such that $p(X_\emptyset =x_\emptyset \mid \Z) = 1$, discarding all empty clique-nodes from the numerator and denominator of \eqref{eq:Z-likelihood-factorization}.

Non-empty non-maximal cliques can either be: i) on the path between two maximal cliques in \(T\), or ii) on a boundary branch of $T$ stemming out of a maximal clique. 

For case i), let $\theta'_{k_1}, \theta'_{k_2}, \dots, \theta'_{k_{n-1}}$ be sub-maximal cliques on the bath between two maximal cliques, $\theta'_{k_0}$ and $\theta'_{k_n}$ in $T$. Let $S = \theta'_{k_0} \cap \theta'_{k_n}$ be the separator representing the edge \((\theta'_{k_0},\theta'_{k_n})\) in $T$. It is straightforward to show that $S \subseteq \theta'_{k_i}$ for all $i=1, \dots, n-1$, otherwise the RIP is violated. There are \(n\) edges for \(n-1\) sub-maximal cliques-nodes in a path between two maximal cliques. For each of the sub-maximal cliques $\theta'_{k_i}\, , i =1, \dots, n-1$, by the RIP, either $\theta'_{k_i} \subseteq \theta'_{k_{i-1}}$, or $\theta'_{k_i} \subseteq \theta'_{k_{i+1}}$, or both. If $\theta'_{k_i} \subseteq \theta'_{k_{i-1}}$ then $p(X_{\theta'_{k_{i}}\cap \theta'_{k_{i-1}}}) = p(X_{\theta'_{k_{i}}})$, thus eliminating the same factor in the numerator of \eqref{eq:Z-likelihood-factorization}. The opposite is similar, when $\theta'_{k_i} \subseteq \theta'_{k_{i+1}}$. This process reduces the path to the single edge \(( \theta'_{k_0},\theta'_{k_n})\) representing \(S\).

For case ii), all sub-maximal clique-nodes on a boundary branch of $T$ stemming out of a maximal clique, say \(\theta'_{k_{0}}\), are contained in \(\theta'_{k_{0}}\). By the RIP, all their edges can be rewired to \(\theta'_{k_{0}}\). Hence, the intersection in the denominator of \eqref{eq:Z-likelihood-factorization} returns the sub-maximal factors as in case i), eliminating them from the numerator.
\end{proof}

The results of Theorem \ref{th:Z-likeli-factor} enables one to use the faster mixing set of (\ref{eq:nei-bound-clique-basic-no-cond}) in the Markov update process without affecting the likelihood of interest. This enables the specification of a multivariate distribution completely in terms of \(\Z\), avoiding the transformation to \(\G\).

\section{Exact sampling conditional on a junction tree}\label{sec:sampling}
Due to the Markovian nature of decomposable graphs, sampling from the proposed framework can be done in few ways. This section illustrates two methods, one based on a sequential procedure with finite number of steps, while the second adapts a Markov update method, where samples are obtained from the stopped process. Nonetheless, both methods overlap in the sampling and embedding of the Poisson process and the assignment of clique-nodes, which is discussed below.

So far, only the location dimension of the Poisson process is considered in the \((r',r)\)-truncation. This risks infinite values for the weight dimension \((\vartheta)\). It is only natural to assume a Poisson process on the \([0,r]\times [0,c]\) cube, where only points with \(\theta < r\) and \(\vartheta < c\) are kept. A standard generative model of nodes and their location embedding can be:

\begin{equation}\label{eq:generating-Nc_Nv}
\begin{aligned}
  &N_v \sim \text{Poisson}(cr),  &N_c &\sim \text{Poisson}(c'r'),\\
  (\theta_i) \mid &N_v \stackrel{{iid}}{\sim} \text{Uniform}[0,r],\quad &(\theta'_k) \mid N_c &\stackrel{{iid}}{\sim} \text{Uniform}[0,r'],\\
  (\vartheta_i) \mid &N_v \stackrel{{iid}}{\sim} \text{Uniform}[0,1],\quad &(\vartheta'_k) \mid N_c &\stackrel{{iid}}{\sim} \text{Uniform}[0,1].\\
\end{aligned}
\end{equation}
where \(N_v\) is the number of nodes and \(N_c\) is the number of clique-nodes.

The iterative sampling of \(T\mid \Z\) is discussed later in Section \ref{sec:sampling-T}. This section only samples a subtree of a given tree by adopting a random walk sampler of edges in \(T\), to avoid the high probability of disjoint components associated with random sampling. The latter could be the case when the tree is known to be finite. The assignment process is then: 
\begin{equation}\label{eq:tree-sampling}
\begin{aligned}
\theta'_1 &\equiv \theta'_{\sigma(1)},\\
\theta'_{n+1}\mid \theta'_1, \dots \theta'_n &\sim \text{Uniform}\Big (\{\theta'_k  \in \Theta': (\theta'_k, \theta'_s) \in \Ep, s\leq n\}\Big),
\end{aligned}
\end{equation}
where \(\sigma(1)\) is a randomly selected clique-node as the root of the sampled tree, and the uniform distribution samples from the neighbouring clique-nodes in \(\Theta'\) to the already assigned ones. 

All the subtree quantities are defined prior to the \((r',r)\)-truncation, thus, we are implicitly assuming that they abide to the condition that \(\theta'_k < r'\).

\subsection{Sequential sampling with finite steps}

Sections \ref{sec:finite-graph-restrictions} and \ref{sec:likel-fact-with} illustrated two perspectives on how the mapping choice between \(\Z\) and \(\G\) affects their projection
relation. Regarding graph embedding in \(\Rp^{2}\), one can relax the bidirectional-injective mapping to surjective-injective by a post-sampling procedure using an identity matrix augmentation. Regarding likelihood factorization, the mapping choice is inconsequential (Theorem \ref{th:Z-likeli-factor}).

An exact sampler with finite number of steps is possible for the surjective-injective mapping case, using the boundary \(\Tbdx{(n)}{i}\) and the neighbouring \(\Tnx{(n)}{i}\) sets defined in \eqref{eq:nei-bound-clique-basic-no-cond}. Otherwise, a finite step sampler can be achieved using the post-sampling procedure discussed in Section \ref{sec:finite-graph-restrictions}.

Because of the dependency induced by \(T\), and as discussed in Section \ref{sec:finite-graph-restrictions}, some nodes might only connect to clique-nodes outside the \((r',r)\)-truncation (non-active). Then, for \(i=1, \dots, N_v\), a node is active within the truncation proportionally to the \([c',r']\) truncation total mass as:

\begin{equation}\label{eq:seq-sampling-node-appears}
\theta_i \textit{ is active } \mid W, c', \vartheta_i  \stackrel{ind}{\sim} \frac{\overline W_1(c', \vartheta_i)}{\overline W_1(\vartheta_i)},  
\end{equation}
where \(\overline W_1(c', \vartheta) = \int_0^{c'}W(x, \vartheta) \d x\).

For each active \(\theta_i\) sample edges as:
\begin{itemize}
\item sample the first edge as
  \begin{equation}\label{eq:sequential-sampling-part1}
    (\theta'_{\pi(k)}, \theta_i) \mid (\vartheta'_k), W \stackrel{ind}{\sim} \frac{W(\vartheta'_{\pi(k)}, \vartheta_i)}{\overline W_1(c',\vartheta_i)}.
  \end{equation}
\item at the \((n+1)\) step, sample edges to neighbouring clique-nodes sequentially as
\begin{equation}\label{eq:sequential-sampling-part2}
\begin{aligned}
\theta'_{\pi(n+1)} \mid (\theta'_{\pi(k)})_{k\leq n} &\sim \text{Uniform}\Big (\Tnx{(n)}{i}\setminus (\theta'_{\pi(k)})_{k\leq n}\Big ),\\
(\theta'_{\pi(n+1)}, \theta_i) \mid  \vartheta'_{\pi(n+1)}, \vartheta_i, W &\sim \text{Bernoulli}\Big ( \frac{W(\vartheta'_{\pi(n+1)}, \vartheta_i)}{\overline W_1(c',\vartheta_i)}\Big ).
\end{aligned}
\end{equation}
\end{itemize}

If it is desired that all non-empty rows of \(\Z^{*}\) represent maximal cliques, then either augment non-empty rows of \(\Z^{*}\) with an identity matrix, or augment non-empty non-maximal rows with single-clique nodes as in the edge-greedy partition. The final bipartite matrix \(\Z^{*}\) can be mapped to a decomposable graph by \eqref{eq:mapping-adj-biadj}.

\subsection{Sampling using a Markov stopped process} \label{sec;markov-update-process}

A sample from a decomposable graph can be obtained from a stopped Markov chain. Such a process is slower in nature than the sequential sampling process discussed in the previous section. In principle, one samples edges uniformly and decides whether they appear at the current step given the current configuration \(\Z\). For the \(n+1\) Markov step, sample edge indices uniformly as
\begin{equation} \label{eq:Sampling-Markov-part1}
\begin{aligned}
  k \mid N_c &\stackrel{iid}{\sim} \text{Uniform}[1, \dots, N_c],\\
  i \mid N_v &\stackrel{iid}{\sim} \text{Uniform}[1, \dots, N_v].\\
\end{aligned}
\end{equation}

Sample the \((\theta'_k, \theta_i)\) edge as
{\small
\begin{equation} \label{eq:Sampling-Markov-part2}
  (\theta'_k, \theta_i) \mid  \vartheta'_k, \vartheta_i, W, T \sim \text{Bernoulli}\Big (W(\vartheta'_k, \vartheta_i) \; \mathbb I\{\theta'_k \in \Tbdx{(n)}{i} \cup \Tnx{(n)}{i} \cup \chi_0^{(n)|i} \} \Big),
\end{equation}}
with
\begin{equation}\label{eq:chi-theta}
\chi_0^{(n)|i}(\theta') = \begin{cases} \theta' & \text{if } |\ver(\Tr{(n)}{i})|=0 \\ \emptyset & \text{otherwise}. \end{cases}
\end{equation}

A realization is then the result of stopping the above iterative process at a random time \(t>0\). Ideally, the stopping time should be chosen after the Markov chain has reached stationarity, such time is referred to as the {\it mixing time} of the Markov chain. The next subsection gives a mixing time result on the Markov stopped process illustrated above. Note that the Markov stopped process can used disregarding the mapping choice, with the proper interpretation.

\subsubsection{Mixing time of the stopped  process}
For a precise definition of the mixing time, let \(\Omega\) be the state space of a Markov chain \((X_t)_{t\geq 0}\) with transition matrix \(P\). Let \(P^t(x,y) = \P(X_t=y \mid X_0=x)\) for \(x,y \in \Omega\), be the probability of the chain reaching state \(y\) in \(t\)-steps given it started at state \(x\). Define the total variation distance \(d(t)\) between the transition matrix \(P^t\), at step \(t\), and the stationary distribution \(\bar p\) as 
\begin{equation}
d(t) := \max_{x \in \Omega} \parallel P^t(x,.) - \bar p \parallel_{TV},
\end{equation} 
where \(\parallel. \parallel_{TV}\) is the total variation norm. Then, the mixing time \(t_{mix}\) is defined as 
\begin{equation}\label{eq:mixing-time}
  t_{mix}:= \min \{t>0 : d(t)<1/4\}.
\end{equation}

Variations of mixing times for other thresholds \(\epsilon \neq 1/4\) exits, though it is shown independently that \(t_{mix}(\epsilon) \leq [\log_2(\epsilon^{-1})]t_{mix}(1/4)\). Therefore, it suffices to work with \eqref{eq:mixing-time}. For an excellent introduction to Markov chain mixing times, refer to the book of \citet{levin2009markov}.

For the proposed sampling method (Sec \ref{sec;markov-update-process}), a unique stationary distribution \(\bar p\) exists, since by construction the chain is irreducible, that is for any two configurations \(x,y \in \Omega\), \(P^t(x,y)>0\) for some \(t\in \N\) \citep[Coro. 1.17, Prop. 1.19]{levin2009markov}. Then, it remains to find a lower bound for \(t_{mix}\). 

A known method to establish lower bounds for mixing times over irreducible Markov chains is by bounding the probability of the first time a {\it coupling} over the chain meets. Given an irreducible Markov chain over a state space \(\Omega\), with transition probability \(P\), a coupling is a process of running two Markov chains \((X_t)_t\) and\((Y_t)_t\), with the same \(P\), though with different starting points.  A coupling meets when the two chains visit a state at the same time and move together at all times after they meet. More precisely, 
\begin{equation}\label{eq:coupling}
\text{if } X_s = Y_s, \text{ then } X_t = Y_t \text{ for } t\geq s.
\end{equation}

\begin{theo}(\citet[Theo. 5.2]{levin2009markov}) Let \(\{(X_t, Y_t)\}\) be a coupling with transition matrix \(P\) satisfying \eqref{eq:coupling}, for which \(X_0 = x\) and \(Y_0=y\). Let \(\tau_{\text{couple}}\) be the first time the chain meets:
\begin{equation}
  \tau_{\text{couple}}:=\min\{t>0 : X_t = Y_t\}.
\end{equation}
Then 
\begin{equation}
d(t) \leq \max_{x, y \in \Omega} \P_{x, y}(\tau_{\text{couple}} >t).
\end{equation}
\label{th:coupling-time}
\end{theo}

An example of a coupling on an \(n\)-node rooted binary tree, is by taking two lazy random walks \((X_t, Y_t)\), started at nodes \(X_0=x, Y_0=y\), where at each step a fair coin decides which chain to move. Then, uniformly move the chosen chain to a neighbouring node, while keeping the other chain fixed. Once the two chains are at the same level from the root node, couple them by moving them further or closer to the root simultaneously. In this case, the first coupling time is less than the {\it commute time} (\(\tau_{0, \partial B}\)), the time a chain commutes from the root to the set of leaves \(\partial B\) and back. By \(\tau_{0, \partial B}\) the coupling would have occurred. 

\begin{myprop}\label{prop:commute-time}(Commute Time Identity \citep[Prop. 10.6]{levin2009markov}) Given a finite tree \(T_n\) with \(n\) nodes, a root node \(x_0\), and a set of leaves \(\partial B\). Let \(\tau_{0,\partial B}\) be the commute time defined as 
\begin{equation}\label{eq:commute-time}
\tau_{0, \partial B} :=\min\{t \geq \tau_{\partial B}: X_t = X_0 =x_0, X_{\tau_{\partial B}}\in \partial B\}, 
\end{equation}
for a random walk \((X_t)_t\) on \(T_n\). Then 
\begin{equation}\label{eq:expected-commute-time}
\E[\tau_{0, \partial B}] = 2(n-1)\sum_{k} \frac{1}{\Gamma_k^{{x_{0}}}},
\end{equation}
where \(\Gamma_k^{x_{0}}\) is the number of nodes at distance \(k\) from the root.
\end{myprop}

\begin{remark} The maximum {\it commute time} is attained for a lazy random walk on a straight line (a path) tree with \(n\) nodes at each side of the root, where \(\E[\tau_{0, \partial B}] = 4n^2\). For a lazy random walk with probability \(p\) that the chain stays at the same configuration, it is easy to see that the expected commute time \eqref{eq:expected-commute-time} becomes \(\E[\tau_{0, \partial B}]/p\).  
\end{remark}
A similar approach could be applied to the proposed sampling scheme of Section \ref{sec;markov-update-process}. First, note that sampling edges for a fixed node \(\theta_i\) depends on the configuration of other nodes. This dependence is enforced by the extra conditions added to \(\Tbd{(n)}{i}\) and \(\Tn{(n)}{i}\) in \eqref{eq:nei-bound-clique-basic} versus \eqref{eq:nei-bound-clique-basic-no-cond}. Following the discussion of Section \ref{sec:finite-graph-restrictions}, we will relax such dependency by decoupling the nodes using the surjective-injective mapping, and the sets in \eqref{eq:nei-bound-clique-basic-no-cond}.

\begin{figure}[!ht]
\captionsetup[subfigure]{labelformat=empty}
  \centering
    \subfloat[]{
    \begin{tikzpicture}[scale=0.65, transform shape, thick]
      \node[circle,draw]  (a) at(0,5){\small $\theta'_1$};
      \node[circle,draw]  (b) at(3,5){\small $\theta'_2$};
      \node[circle,draw]  (c) at(1.5,3){\small $\theta'_3$};
      \node[circle,draw]  (d) at(0,1){\small $\theta'_4$};
      \node[circle,draw]  (e) at(3,1){\small $\theta'_5$};
      \draw[-latex] (b) to[bend right=10] node[above,rotate=0] {\small $W_1$} (a);
      \draw[-latex] (a) to[bend right=10] node[below,rotate=0] {\small $W_2$} (b);

      \draw[-latex] (a) to[bend right=10] node[below,rotate=-45] {\small $W_3$} (c);
      \draw[-latex] (c) to[bend right=10] node[above,rotate=-45] {\small $W_1$} (a);

      \draw[-latex] (c) to[bend right=10] node[below,rotate=-45] {\small $W_5$} (e);
      \draw[-latex] (e) to[bend right=10] node[above,rotate=-45] {\small $W_3$} (c);

      \draw[-latex] (c) to[bend right=10] node[above,rotate=45] {\small $W_4$} (d);
      \draw[-latex] (d) to[bend right=10] node[below,rotate=45] {\small $W_3$} (c);
    \end{tikzpicture}}\quad \quad 
    \subfloat[]{
    \begin{tikzpicture}[scale=0.65, transform shape, thick]
      \node[circle,draw]  (a) at(0,5){\small $\theta'_1$};
      \node[circle,draw]  (b) at(3,5){\small $\theta'_2$};
      \node[circle,draw]  (c) at(1.5,3){\small $\theta'_3$};
      \node[circle,draw]  (d) at(0,1){\small $\theta'_4$};
      \node[circle,draw]  (e) at(3,1){\small $\theta'_5$};

      \draw (b) to[bend right=0] node[above,rotate=0] {\small $W_*$} (a);
      \draw (a) to[bend right=0] node[below,rotate=-45] {\small $W_*$} (c);
      \draw (e) to[bend right=0] node[above,rotate=-45] {\small $W_*$} (c);
      \draw (c) to[bend right=0] node[above,rotate=45] {\small $W_*$} (d);
    \end{tikzpicture}}
  \caption{A 5-node tree: left is the original directed weighted tree where $W_k:= W(\vartheta'_k, \vartheta_i)$ for a random $\vartheta_i$, right is the undirected tree by expectation where $W_* = \E(W)$.}
\label{fig:weighted-tree}
 \end{figure}
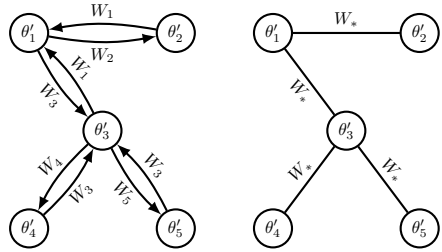

The objective of breaking down the dependency between nodes is to reduce the problem of studying the mixing time on the whole graph to studying it on each node independently. In this case, the process in \eqref{eq:Sampling-Markov-part2} does not map directly to a random walk process, where we can apply the commute time identity. For three reasons: i) for each node \(\theta_i\), the edges of the junction tree are directional and weighted by \(W(\,.\,, \vartheta_i)\); ii)  the variable \(\chi_0^{(n)|i}\) in \eqref{eq:chi-theta} acts like a transporting hub to a random clique-node whenever the random walk returns to the starting position; iii) the commute time in \ref{prop:commute-time} depends on a root node, that is not a property of the proposed sampling method. Nonetheless, all three reasons can be handled. In i), for a non-atomic \(W\) a uniform expected weight of
\begin{equation} \label{eq:expected-weight}
E[W] = \iint_{\Rp^2}W(x,y)\d x \d y,
\end{equation}
can be used. It is attained by a direct application of the Mapping theorem of \citep{kingman1993poisson}, as in Figure \ref{fig:weighted-tree}. For ii), the transport hub property only speeds up the commute time, thus an upper bound is still the commute time of \eqref{eq:commute-time}. For iii), \(\sum_{k}1/\Gamma_{k}\) is smallest when the designated root node is the centre of the tree, where each side is symmetric. It becomes larger as the designated root node moves away from the centre, with the maximum of \(L^{\max}/2\), half the maximum distance between two leaf nodes.

\begin{mylemma}\label{lem:mixing-time} For the Markov update process of Section \ref{sec;markov-update-process}, given a connected tree with \(N_c\) clique-nodes, the lower bound of the expected mixing time for an arbitrary node is
\begin{equation}\label{eq:mixing-time-each-node}
t_{\text mix}\geq \frac{8N_c}{\iint_{\Rp^2} W(x, y)\d x\d y} \frac{L^{\max}}{2} \geq  \frac{8N_c}{\iint_{\Rp^2} W(x, y)\d x\d y} \sum_{k=1}^{N_c}\frac{1}{\Gamma_k}, 
\end{equation}
where \(\Gamma_k \) is number of clique-nodes at distance \(k\) form a root clique-node selected randomly from the non-leaf nodes of the tree, and \(L^{\max}\) is the maximum distance between two leaf clique-nodes. When nodes are sampled independently, then \eqref{eq:mixing-time-each-node} is the global mixing time achieved by parallel sampling.
\end{mylemma}
The proof follows directly from Theorem \ref{th:coupling-time} and Proposition \ref{prop:commute-time} by a lazy random walk with probability as in \eqref{eq:expected-weight}.

\section{Edge updates on a junction tree}\label{sec:sampling-T}

Section \ref{sec:GenerModel} proposed a model for decomposable random graphs by conditioning on a fixed junction tree, where graph edges are formed conditionally through a Markov process, as shown in \eqref{eq:Update-step-simple} and Section \ref{sec:sampling}. The iterative counterpart is a model to sample \(T \mid \Z\).

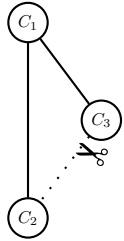
\begin{figure}[!ht]
  \centering
    \begin{tikzpicture}[scale=0.65, transform shape, thick]
      \node[circle,draw]  (a) at(0,5){\small $C_1$};
      \node[circle,draw]  (c) at(1.5,3){\small $C_3$};
      \node[circle,draw]  (d) at(0,1){\small $C_2$};
      \draw (a)--(c);
      \draw (d)--(a);
      \draw[loosely dotted] (c)--(d);
      \node[rotate=-20] (scissors) at (1.3,2.3) {\Huge \Leftscissors};
    \end{tikzpicture}
\caption{Moving along the bipartite graph of Figure \ref{fig:ex_bipartite}, from junction tree $T_1$ to $T_2$, through severing and reconnecting the edge $(C_2, C_3)$ (dotted lines) to $(C_2, C_1)$.}
  \label{fig:example-cutting-single-edge-junction-tree}
 \end{figure}

Despite the non-uniqueness of junction trees and the POSs, \citet{lauritzen1996} has showed that the set of minimal separators, edges of a junction tree, is unique with varying multiplicity for each separator. The separator multiplicity relates to the number of ways its corresponding edge can be formed, and thus the number of trees that are a unit distance, or a single move, away. Therefore, for two adjacent cliques \(\theta'_{k}\) and \(\theta'_{s}\) in some tree \(T\), if \(\G(\theta'_k)\cap \G(\theta'_s) \subset \G(\theta'_m)\), for a third clique \(\theta'_{m}\), then one can alter the edge \((\theta'_k, \theta'_s)\) by severing it on one side and reconnecting it to \(\theta'_m\). For example, in Figure \ref{fig:ex_bipartite}, moving from the junction tree \(T_1\) in the Subfigure \ref{fig:ex_bipartite-B} to \(T_2\), requires the severing of edge \((C_2, C_3)\) from the \(C_3\) side and reconnecting it to \(C_1\), as shown in the Figure \ref{fig:example-cutting-single-edge-junction-tree}. The separating nodes between \(C_2\) and \(C_3\) are \(\G(C_2) \cap \G(C_2) = \{\text{CD}\}\) and are contained in the clique \(C_1 = \{\text{ABCD}\}\). 

 The set of clique-nodes a severed edge can reconnect to is the same set of clique-nodes that satisfy the running intersection property of the POSs, introduced in \eqref{eq:perfectorder}. To formalize this notion, for some tree \(T = (\Theta', \Ep)\) and edge \((\theta'_k, \theta'_s) \in \Ep\), let \(\JT{}{k}{s}\) be the set of cliques that satisfy the RIP when the edge is severed at the \(\theta'_{s}\)'s side, as
\begin{equation}\label{eq:junction-tree-set-possible-cliques}
  \JT{}{k}{s} = \Big \{\theta'_m :(\theta'_k, \theta'_s) \in \Ep, \theta'_k\cap \theta'_s \subset \theta'_m, (\theta'_{k'}, \theta'_{s}) \in \theta'_{s}\sim \theta'_{m} \Big\}.
\end{equation}

The notation \((\theta'_{k'}, \theta'_{s}) \in \theta'_{s}\sim \theta'_{m} \) indicates that \((\theta'_{k'}, \theta'_{s})\) is in the path between \(\theta'_{s}\) and \(\theta'_{m}\) in \(T\), where \(\theta'_s \in \JT{}{k}{s}\). Let \(\JE{}{k}{s}{m}=1\) be the indicator that the edge \((\theta'_k, \theta'_s)\) is replaced by \((\theta'_k,\theta'_{m})\). Using a uniform prior, the probability of such move is

\begin{equation}\label{eq:junction-tree-edge-update-uniform}
\P(\JE{}{k}{s}{m}=1 \mid \Z, T) = 
\begin{cases}
\frac{1}{|\JT{}{k}{s}|} & \text{if } \theta'_m \in \JT{}{k}{s} \\
0                         & \text{otherwise}.
\end{cases}
\end{equation}

A weighted version can also be formed. For example, when larger cliques are favoured over smaller ones, the update distribution can take the form
\begin{equation}\label{eq:junction-tree-edge-update-weighted}
\P(\JE{}{k}{s}{m}=1 \mid \Z, T) = 
\begin{cases}
\frac{\deg(\theta'_m, \Z)}{\sum_{x \in \JT{}{k}{s}}\deg(x, \Z)} & \text{if } \theta'_m \in \JT{}{k}{s} \\
0                         & \text{otherwise}, 
\end{cases}
\end{equation}

To combine the results with the ones of Section \ref{sec:sampling}, an iterative sampling of the decomposable graph and the tree is:

\begin{enumerate}[(i)]
\item generate \(N_v, (\theta_i), N_c\) and \((\theta'_k)\) as in \eqref{eq:generating-Nc_Nv};
\item sample an initial tree by the random assignment process in \eqref{eq:tree-sampling};
\item at the \(n\)-th Markov step:
  \begin{itemize}
    \item sample \(\Z^{(n+1)} \mid T^{(n)}\) according to samplers of Section \ref{sec:sampling};
      \item sample \(T^{(n+1)} \mid \Z^{(n+1)}\) according to \eqref{eq:junction-tree-edge-update-uniform}, or its weighted version.
  \end{itemize}
\end{enumerate}

\section{Model properties: expected number of cliques} \label{sec:expected-num-cliques-per-node}
Lemma \ref{lem:degree-dist-regular-biadjacency} defined the degree function of a regular bipartite measure. The set \(\Pi_r\cup\{(\theta, \vartheta)\}\) is used to properly define the conditioning on the null set \((\theta, \vartheta) \in \Pi_r\), by application of the Slivnyak-Mecke theorem.

The expression of the degree function in \eqref{eq:degree-dist-bipartite} does not hold for the proposed decomposable random graphs model of Definition \ref{def:decomposable-graph-definition}. First, because the set \(\Pi_{r'}\) of clique-nodes carries a dependency structure based on \(T\). Second, the Markovian nature of the process restricts the node-clique (dis)connections to the set of boundary and neighbouring clique-nodes by means of \eqref{eq:nei-bound-clique-basic-no-cond} or \eqref{eq:nei-bound-clique-basic}. Nonetheless, with the product of distinct Poisson process identity \citep[Chap. 3.1 Eq. 3.14]{kingman1993poisson}, an analogous degree function can be defined. Let \(\Pi_x\) be a rate \(x\) homogeneous Poisson process, and following the notations of Lemma \ref{lem:degree-dist-regular-biadjacency}.
\begin{mylemma} For a tree-dependent bipartite process \(\Z\) of Definition \ref{def:decomposable-graph-definition}, assuming a surjective projection into the space of decomposable graphs using \eqref{eq:nei-bound-clique-basic-no-cond}, with a non-random \(W:\Rp^2 \mapsto [0,1]\), the degree of a node \((\theta, \vartheta) \in \Rp^2\), given an initial connection to clique node \((\theta'_0, \vartheta'_0) \in \Rp^2\) is 
  \begin{equation}\label{eq:degree-function-decomposable}
    \begin{aligned}
      \deg(\theta, \theta'_{0}, \Z_{r',r})&:=\deg\big(\theta, \Pi_r\cup\{(\theta,\vartheta)\}, \Pi_{r'}\cup\{(\theta',\vartheta'_0)\}, (U_{ki}), T\big)\\
      &:=  \sum_{k=0}^\infty \Big [\sum_{\substack{y \in \Pi_{r'}\\ y \in     \widetilde{\Gamma}^{\theta'_0}_k}}\prod_{ s \in \mathcal{P}(\theta'_0 \rightarrow y)} \mathbb{I}\{U_{k(s)i(\theta)} \leq W(s, \vartheta)\} \Big ],
    \end{aligned}
  \end{equation}    
where \((U_{ki})\) is a 2-array of uniform[0,1] random variables, with \(k(x)\) and \(i(y)\) being abbreviations for the index of \(x\) and \(y\). \(\widetilde {\Gamma}^{\theta'_0}_k\) is the set of clique-nodes in \(\Pi_{r'}\) at distance \(k\) from \(\theta'_0\), where \(\widetilde{\Gamma}^{\theta'_0}_0= \theta'_0\). \(\mathcal{P}(\theta'_0 \rightarrow y)\) is the set of clique-nodes on the path from \(\theta'_0\) to \(y\). Moreover, the expectation of \eqref{eq:degree-function-decomposable} is
\begin{equation}\label{eq:exp-degree-function-decomposable}
  \E[ \deg\big(\theta,\theta'_0,\Z_{r',r})] = \sum_{k=0}^\infty \Gamma^{\theta'_0}_k\big ( r'\overline W_1(\vartheta)\big )^{k+1},
\end{equation}
where \(\Gamma^{\theta'_0}_k =|\widetilde{\Gamma}^{\theta'_0}_k|\). For a random \(W\), the results can be seen by conditioning.
\label{lem:expected-degree-decomposable}
\end{mylemma}
\begin{proof}
By invoking the Slivnyak-Mecke theorem twice for the event \((\theta, \vartheta) \in \Pi_{r}\) and \((\theta'_{0},\vartheta'_0) \in \Pi_{r}\). Let  \(y_0, y_1, y_2, \dots, y_n \in \Pi_{r'}\) be a series of locations of clique-nodes on the path from \(y_0\) to \(y_n\), where \(y_s\) is at distance \(s\) from \(y_0\).  By the Markovian nature of tree-dependent bipartite processes, for an edge \((y_s, \theta)\) to form with probability larger than 0, \(y_s\) must be a neighbouring clique-node to \(\Tr{}{(\theta)}\), that is  \(y_s \in \Tnx{}{\theta} \) of \eqref{eq:nei-bound-clique-basic-no-cond}, implying \(y_0, y_1, \dots, y_{s-1} \in \Tr{}{\theta}\). Thus, the event that \((y_s, \theta) \in \Z\) amounts to 
\begin{equation}
\prod_{j=0}^{s}\mathbb{I}\{U_{k(y_j), i(\theta)}\leq W(y_j, \vartheta)\}.
\end{equation}

By uniqueness of paths in trees and the ordering of clique-nodes by distance \(k\) in the set \(\widetilde{\Gamma}^{\theta'_0}_k\) from an assumed initial point \(\theta'_0\), \eqref{eq:degree-function-decomposable} is obtained. For \eqref{eq:exp-degree-function-decomposable}, the Poisson process identity \citep[Chap. 3.1 Eq. 3.14]{kingman1993poisson} is directly applicable, since the set \(\widetilde{\Gamma}^{\theta'_0}_k\) contains distinct points of \(\Pi_{r'}\). Thus, for each \(y \in \widetilde{\Gamma}^{\theta'_0}_k\) the inner sum of \eqref{eq:degree-function-decomposable} becomes
\begin{equation}
\prod_{ s \in \mathcal{P}(\theta'_0 \rightarrow y)}\E\Big [\sum_{s \in \Pi_{r'}} \mathbb{I}\{U_{k(s)i(\theta)} \leq W(s, \vartheta)\} \Big ] = \prod_{ s \in \mathcal{P}(\theta'_0 \rightarrow y)} \int_{\Rp}W(s, \vartheta)r' \d s.
\end{equation}

For \(y \in \widetilde{\Gamma}^{\theta'_0}_k\), the path length is \(|\mathcal{P}(\theta'_0 \rightarrow y)| = k+1\), and \eqref{eq:exp-degree-function-decomposable} follows.
\end{proof}

The degree expectation \eqref{eq:exp-degree-function-decomposable}, depends on the initial clique-node \((\theta'_0, \vartheta'_{0})\) through the number of clique-nodes at a certain distance from it, as indicated by \((\Gamma_k)_k\). Therefore, for certain tree structures, for example \(d\)-regular trees, the sizes of \((\Gamma_k)_k\) are explicit functions of tree degree distribution. In such cases, a more explicit characterization of \eqref{eq:exp-degree-function-decomposable} is achievable. The following corollary gives a compact expression of the expected cliques per node (clique-degree) for \(d\)-regular trees, where \(d\geq 3\). We refer to \(d\)-regular trees as trees where all nodes are of degree \(d\), thus, a binary tree is also a \(3\)-regular tree.

\begin{mycoro} \label{cor:expectation-per-level} Let \(T\) be a \(d\)-regular tree with \(d\geq 3\), a root clique-node \({(\theta'_{0},\vartheta'_{0})}\), and \(L\in \N\) levels. Such that each clique-node has degree \(d\), except for leaf nodes with degree 1. A clique-node \({{\theta'}}^{\ell}_{k}\) is said to be at level \(\ell \in \{0, 1, \dots L\}\) if the distance between \({\theta'}_0\) and \({\theta'}_{k}^\ell\) is \(\ell\). For \(T\)-dependent bipartite graph \(\Z\), with junction tree \(T\), the expected number of clique connections (clique-degree) for a node \(\theta\) with an initial connection to clique-node \({\theta'}_{k}^\ell\) is
  \begin{equation}\label{eq:expected-degree-d-regular-tree}
    \begin{aligned}
      \E[\deg\big(\theta,{\theta'}_{k}^\ell,\Z_{r',r})\mid {\theta'}_{k}^\ell \in \ell]= \zeta &+ \zeta^2 d \frac{(\dd \zeta)^{L-\ell}-1}{\dd\zeta -1}\\
      &+ \zeta^2(\dd\zeta)^{L-\ell}(\dd\zeta+1) \frac{(\dd \zeta^2)^\ell -1}{\dd \zeta^2 -1},   
    \end{aligned}
  \end{equation}
where \(\zeta = r'\overline W_1(\vartheta)\) and \(\dd = (d-1)\). Such that, for an initial starting point at the root \({\theta'}_0\), the expected value simplifies to
\(\zeta + \zeta^2 d \big (\dd^L\zeta^{L}-1\big )/\big (\dd\zeta -1\big )\). For \({\theta'}_{k}^L\), a node in level \(L\), it is \(\zeta + \zeta^2 (\dd\zeta + 1) \big ( \dd^L\zeta^{2L}-1\big )/\big ( \dd\zeta^2 -1 \big )\).
\end{mycoro}

\begin{proof}
  With some simple algebra, few properties of \(d\)-regular trees are accessible, for example, the number of clique-nodes at distance \(0<k<L\) from the root \({\theta'}_0\) is \(\Gamma_k^{{\theta'}_0} = d(d-1)^{k-1}\), where \(\Gamma_k^{{\theta'}_0}=0\) for \(k>L\). Other properties require more combinatorial work, the interest is in defining the more general expression of \(\Gamma_k^{\ell}\), which is the number of clique-nodes at distance \(k\) from clique-nodes at level \(\ell \in \{0,1,2, \dots, L\}\). The following is a list of simple properties of \(d\)-regular trees with root node \({\theta'}_{0}\), which will come useful in defining \(\Gamma_k^{\ell}\):

\begin{enumerate}[(a)]
\item for a fixed \(\ell\), the maximum distance is \(\max_{k}\{\Gamma_k^\ell >0\} = L + \ell \), that is \(2L\) for \(\ell=L\);
\item for a tree with \(n\) nodes, \(\sum_{k\geq 0} \Gamma_k^{\ell}=n\), for all \(\ell\), where \(\Gamma_0^{\ell}=1\);
\item \(\Gamma_k^{\ell}=\Gamma_k^{{\theta'}_0}\) for all \(k \leq L-\ell\);
\item by the geometric sum, the number of nodes in \(T\) in terms of \(d\) and \(L\) is
\begin{equation}\label{eq:number-of-nodes-in-T}
|\ver(T)| = \sum_{k=0}^{L}\Gamma_k^{{\theta'}_0} = 1+ d\frac{(d-1)^L-1}{d-2} = \frac{d\dd^L -2}{d-2}.
\end{equation}
\end{enumerate}

Properties (b) and (c) show that for distances larger than \(L-\ell\), the standard distribution rule of \(\Gamma_k^{\theta'_0} = d(d-1)^{k-1}\) does not apply. By combinatorial counting and induction, Table \ref{tb:nodes-at-distance-from-level} summarizes the general expression for \(\Gamma_k^\ell\) for different values of \(\ell\), where \(\dd = (d-1)\) and $\lfloor x\rfloor$ is the floor operator. The values in the top left corner of Table \ref{tb:nodes-at-distance-from-level}, bordered by the ladder shape, corresponds to property (c) above. Moreover, for each row the values under the ladder come in pairs as a result of floor operator, except the first (i.e. 1 for row \(L\), \(d\) for row \(L-1\)) and the last value (\(\dd^L\)). Therefore, the total number of clique-nodes at all distances for clique-nodes in level \(\ell\) is 
\begin{equation}\label{eq:sum-of-row-for-each-level}
\begin{aligned}
\sum_{k=1}^{2L}\Gamma_k^{\ell} &= \underbrace{\sum_{k=1}^{L-\ell}d (d-1)^{k-1}}_{\text{part above ladder in \ref{tb:nodes-at-distance-from-level}}} + \underbrace{\sum_{k=0}^{\ell} (d-1)^{L-k} \big [ \delta_{(0, L]}(k) +\delta_{(0, L]}(k+1) \big ]}_{\text{part under ladder}} \\
&=\sum_{k=1}^{L-\ell}d (d-1)^{k-1}+ \underbrace{\sum_{k=0}^{\ell} 2(d-1)^{L-k} - \big [ (d-1)^L + (d-1)^{L-\ell}\big]}_{\text{with correction for first and last values}}, 
\end{aligned}
\end{equation}
where \(\delta_{(0, L]}(s) = 1\) if \(0<s \leq L\).
\setlength\tabcolsep{2.5pt}
\begin{table}[!ht]
\centering
\caption{A summary table of the number of clique-nodes at distance $k$ from clique-nodes at level $\ell\leq L$ for a $d$-regular tree with $L$ levels, where $\lfloor x\rfloor$ is the floor operator and $\dd =(d-1)$.}
\label{tb:nodes-at-distance-from-level}
{\footnotesize
\begin{tabular}{|l |ccccccccccc| }
\hline
  & \multicolumn{11}{c|}{Number of clique-nodes at distance $k$} \\
\hline
   $\ell$& 1     & 2        & 3          & \dots        & \(L-1\)        & \(L\)          & \(L+1\)      & \dots& \(2L-2\)& \(2L-1\) & \(2L\) \\
\hline
 0 & \(d\) & \(d\dd\) & \(d\dd^2\) & \dots  & \(d\dd^{L-2}\) &\multicolumn{1}{c|}{\(d\dd^{L-1}\)} & 0                       & \dots     &0 & 0 &0  \\ \cline{7-7} 
1  & \(d\) & \(d\dd\) & \(d\dd^2\) & \dots  & \multicolumn{1}{c|}{\(d\dd^{L-2}\)} & \(\dd^{L-1}\)  & \(\dd^{L}\)             & \dots      &0  &0 &0 \\ \cline{6-6}
2  & \(d\) & \(d\dd\) & \(d\dd^2\) & \multicolumn{1}{c|}{\dots}  & \(\dd^{L-2}\)  & \(\dd^{L-1}\)  & \(\dd^{L-1}\) & \dots       &0 & 0 &0 \\
\vdots & \vdots & \vdots & \vdots & \vdots & \vdots  & \vdots & \vdots & \vdots  & \vdots & \vdots & \vdots \\\cline{4-4}
\(L-2\)           & d & \multicolumn{1}{c|}{\(d\dd\)}&\(\dd^2\)& \dots   &\(\dd^{L-\lfloor (L)/2\rfloor}\)  &\(\dd^{L-\lfloor (L-1)/2 \rfloor}\) &  \(\dd^{L-\lfloor (L-2)/2 \rfloor}\)  &    \dots & \(\dd^{L}\) &0&0 \\\cline{3-3}
\(L-1\)           & \multicolumn{1}{c|}{\(d\)} & \(\dd\)&\(\dd^2\)& \dots  &\(\dd^{L-\lfloor (L+1)/2\rfloor}\)  &\(\dd^{L-\lfloor L/2 \rfloor}\) &  \(\dd^{L-\lfloor (L-1)/2 \rfloor}\)   &\dots &\(\dd^{L-1}\) &  \(\dd^{L}\) & 0  \\\cline{2-2}
\(L\)             & 1 & \(\dd\)&\(\dd\)& \dots &\(\dd^{L-\lfloor (L+2)/2\rfloor}\)  &\(\dd^{L-\lfloor (L+1)/2 \rfloor}\) &  \(\dd^{L-\lfloor L/2 \rfloor}\) &     \dots &\(\dd^{L-1}\)&\(\dd^{L-1}\) & \(\dd^L\)  \\
\hline
\end{tabular}
}
\end{table}

To arrive at \eqref{eq:expected-degree-d-regular-tree}, using \eqref{eq:exp-degree-function-decomposable}, with \(\zeta = r'\overline W_1(\vartheta)\) and \(\dd = (d-1)\), the logic used in \eqref{eq:sum-of-row-for-each-level} gives 
{\small
\begin{equation}
\begin{aligned}
&\sum_{k=0}^\infty \Gamma^{\ell}_k \zeta^{k+1} = \zeta + \sum_{k=1}^{2L}\Gamma_k^{\ell}\zeta^{k+1} \\
&= \zeta + \sum_{k=1}^{L-\ell}d\dd^{k-1}\zeta^{k+1} + \sum_{k=0}^\ell \dd^{L-k}\big [ \zeta^{L+\ell-2k+2}\delta_{(0, L]}(k) +\zeta^{L+\ell-2k+1}\delta_{(0, L]}(k+1) \big ] \\ 
&= \zeta + \sum_{k=1}^{L-\ell}d\dd^{k-1}\zeta^{k+1} + \sum_{k=0}^\ell \dd^{L-k}\zeta^{L+\ell-2k+1}(\zeta + 1) - \big [\dd^L\zeta^{L+\ell +2} + \dd^{L-\ell}\zeta^{L-\ell+1}\big].
\end{aligned}
\end{equation}
}
The form in \eqref{eq:expected-degree-d-regular-tree} follows directly from multiple applications of geometric series sum and simplification.
\end{proof}

The simplified expectation form in Corollary \ref{cor:expectation-per-level} is still restrictive, though it required combinatorial work. The following corollary generalizes the result by extending it for an arbitrary initial point of any level. 

\begin{mycoro} \label{cor:expectation-arbitrary} Following the settings of Corollary \ref{cor:expectation-per-level}, the expected clique-degree of a node \(\theta\), for an arbitrary initial starting point, is 

\begin{equation}
\begin{aligned}
  \E[\deg\big(\theta,\Z_{r',r})] = \frac{d-2}{d\dd^L-2}&\Bigg [ \zeta + d\zeta^2 \frac{(\dd\zeta)^L-1}{\dd\zeta -1} - \zeta d\frac{\zeta +1}{\dd\zeta-1}\frac{\dd^L-1}{\dd-1} \\
&+ \zeta^2 d \dd^{L} \frac{\zeta^L-1}{\dd\zeta^2 -1}\frac{\zeta+1}{\dd\zeta-1} \\ 
&+ \zeta^3d(\dd\zeta)^L \frac{\dd\zeta+1}{\dd\zeta^2-1}\frac{(\dd\zeta)^L-1}{\dd\zeta-1}\Bigg]. 
\end{aligned}
\end{equation}
\end{mycoro}

The proof is directly obtainable by linearity of expectation on disjoint domains, where the probability that the initial point is in level \(\ell\) is \(\Gamma^{{\theta'}_0}_\ell/|\ver(T)|\).

Corollary \ref{cor:expectation-per-level} and \ref{cor:expectation-arbitrary} illustrate the case of \(d\)-regular trees, where \(d\geq 3\) with \(d=3\) being the binary tree. For the case of a path junction tree, where \(d=2\), a very similar and simpler result is possible.

\begin{mycoro} \label{cor:expectation-path-tree} Let \(T\) be \(2\)-regular tree with root clique-node \({\theta'}_0\) and \(L\in \N\) levels. Then, \(T\)-dependent bipartite graph \(\Z\), the expected clique-degree of a node \(\theta\) with an initial clique-node \({\theta'}^\ell\) in level \(\ell \in \{0, 1, \dots, L\}\) is
  \begin{equation}
    \label{eq:expected-degree-per-level-binary-tree}
    \E[\deg\big(\theta,{\theta'}^\ell,\Z_{r',r}) \mid {\theta'}^\ell \in \ell] = \zeta + 2\zeta^2\frac{\zeta^{L-\ell}-1}{\zeta-1} + \zeta^{L-\ell+1}\frac{\zeta^{2\ell}-1}{\zeta-1},
  \end{equation}
where \(\zeta = r'\overline W_1(\vartheta)\). For an arbitrary initial point, the expectation becomes
\begin{equation}
    \E[\deg\big(\theta,\Z_{r',r})] = \frac{\zeta}{2L+1}\Bigg [1 - 2L \frac{\zeta+1}{\zeta-1}+2(\zeta^{L+1}+\zeta^2 +\zeta -1)\frac{\zeta^L-1}{(\zeta-1)^2} \Bigg].
\end{equation}
\end{mycoro}

The proof of Corollary \ref{cor:expectation-path-tree} follows the same derivation method of \ref{cor:expectation-per-level} and \ref{cor:expectation-arbitrary}, thus it is omitted. We now illustrate few expectation examples for small \(d\)-regular trees. 

\begin{example} According to Corollary \ref{cor:expectation-arbitrary}, for the binary junction tree with \(L=2\), the expected clique-degree of an arbitrary node is 
  \begin{equation}
    \label{eq:expected-degree-binary-tree-10-nodes}
 \frac{\zeta}{5}(12\zeta^4+12\zeta^3+12\zeta^2+9\zeta+5).
  \end{equation}

For \(L=3\) it is 
\begin{equation}
\frac{\zeta}{11}(48\zeta^6+48\zeta^5+48\zeta^4+36\zeta^3+30\zeta^2+21\zeta+11).
\end{equation}
\end{example}

\begin{example} By Corollary \ref{cor:expectation-path-tree}, for a path junction tree with \(L=2\) levels (5 clique-nodes), the expected clique-degree of an arbitrary node is
  \begin{equation}
  \frac{\zeta}{5}(2\zeta^3+6\zeta^2+10\zeta+7).
  \end{equation}

For \(L=3\), 7 clique-nodes, it is 
\begin{equation}
 \frac{\zeta}{7}(2\zeta^5+4\zeta^4+8\zeta^3+12\zeta^2+14\zeta+9).
\end{equation}
\end{example}

\section{Examples}\label{sec:examples}
This section aims to showcase a few practical examples of decomposable graphs under different choices of \(W\) and the \((r',r)\)-truncation, where the unit rate Poisson process is used as a sampling mechanism.

In some recent work, the \(W\) is treated as a limit object of a series of graph realizations \citep{gao2015rate,wolfe2013nonparametric}, in others, \(W\) is treated as a deterministic function of distributional family \citep{caron2014sparse}. This section follows the latter by letting \(W\) be a deterministic function of some known parametric distributions, and the interest is in estimating the distributional parameters given a realization.

Sampling from parametric distributions is usually done through the right-continuous inverse of the distributional CDF by means of a uniform random variable. There is a direct link between unit rate Poisson processes and uniform random variables, that can be shown in few ways. For example, using distributional equality the unit rate Poisson observations \((\vartheta_i)\) can be ordered such that \(\vartheta_{(1)}< \vartheta_{(2)}< \dots\), then \(\vartheta_{(i+1)}- \vartheta_{(i)} \sim \text{Exponential}(1)\), as the inter-arrival times between events. As a result, \(\exp(-(\vartheta_{(i+1)}- \vartheta_{(i)})) \sim\) Uniform\([0,1]\).

The tree-dependent bipartite representation of decomposable graphs results in a simple expression for the conditional joint distribution, however, the conditioning choice is important, as shown in the following subsection.

\subsection{On the joint distribution of a realization}\label{sec:joint-distr-real}
The Markov nature of decomposable graphs forces nodes to establish their clique connections in \(\Z\) via a path over \(T\). For example, a node \(\theta_{i}\) initially connects to the clique \(\theta'_{\sigma(1)}\); attempts unsuccessfully to connect to neighbouring cliques-nodes of \(\theta'_{\sigma(1)}\) in \(T\); with a successful connection to \(\theta'_{\sigma(2)}\); \(\theta_{i}\) attempts the neighbours of \(\theta'_{\sigma(2)}\) that are not yet attempted, and so on. This results in \(\Tr{}{i}\), which defines the successful connection path of \(\theta_{i}\), the unsuccessful attempts are defined by \(\Tn{}{i}\).

Let \(\Z^{\star}\) be a fixed size matrix, as in the discrete case, then by disregarding the initial starting clique for node \(\theta_{i}\), by conditioning on all other connections and a tree \(T\), the joint distribution of \(\z_{.i}^{\star}\), the \(i\)-th column of \(\Z^{\star}\), can be defined as
{\small
\begin{equation}\label{eq:joint-distribution-general}
  \begin{aligned}
    \P(\z^{\star}_{.i} \mid \Z^{\star}_{-(.i)}, T) &= \Bigg \{ \prod_{\theta' \in \ver(\Tr{}{i})} \P(z^{\star}_{k(\theta')i}=1) \Bigg \}\Bigg \{ \prod_{\theta' \in \ver(\Tn{}{i})} \P(z^{\star}_{k(\theta')i}=0 )\Bigg \}\\
   & = \Bigg \{ \prod_{\theta' \in \ver(\Tr{}{i})} W(\vartheta'_{k(\theta')}, \vartheta_{i}) \Bigg \}\Bigg \{ \prod_{\theta' \in \ver(\Tn{}{i})} 1-  W(\vartheta'_{k(\theta')}, \vartheta_{i})\Bigg \},
    \end{aligned}
\end{equation}}
where \(k(\theta')\) is the index of \(\theta'\) and \(\Z^{\star}_{-(.i)}\) is \(\Z^{\star}\) excluding the \(i\)-th column.

Therefore, for an observed decomposable graph \(\G\), with \(N_{c}\) maximal cliques, having \(\Z^{\star}\) as its biadjacency matrix, define the following neighbourhood indicator as
\begin{equation}\label{eq:posterior-dist-delta-quantity}
\deltanei_{ki}  = \left\{
 \begin{array}{l l}
    1 & \quad \text{if } \theta'_k  \in \Tn{}{i}\\
    0 & \quad \text{otherwise},
  \end{array} \right.  
\end{equation}
where \(\Tn{}{i}\) as in \eqref{eq:nei-bound-clique-basic}. Then \eqref{eq:joint-distribution-general} simplifies to
{\small
\begin{equation}\label{eq:joint-distribution-general-W}
  \begin{aligned}
    \P(\z^{\star}_{.i} \mid \Z^{\star}_{-(.k)}, T) = \prod_{k=1}^{N_{c}}\Bigg \{ W(\vartheta'_{k(\theta')}, \vartheta_{i}) \Bigg \}^{z_{ki}}\Bigg \{ 1-  W(\vartheta'_{k(\theta')}, \vartheta_{i})\Bigg \}^{(1-z_{ki})\deltanei_{ki}}.
    \end{aligned}
  \end{equation}
}

The dependence on all other node-clique connections \(\Z^{\star}_{-(.i)}\) in \eqref{eq:joint-distribution-general} is a direct result of using the quantity \(\Tn{}{i}\), which includes clique-nodes that are neighbouring to \(\Tr{}{i}\) that do not cause a maximal clique to be sub-maximal (Eq. (\ref{eq:nei-bound-clique-basic})). Not all columns of \(\Z^{\star}\) exhibit such dependency, nonetheless, the conditions causing \(\z^{\star}_{.i}\) to depend on \(\Z^{\star}_{-(.i)}\) in \eqref{eq:nei-bound-clique-basic} are only an artifact of the proposed sampling process in (\ref{eq:Update-step-simple}), to force every non-empty node of a finite \(T\) to be maximal at each Markov step. Proposition \ref{prop:decom-graph-sub-maximal-treatment} and the discussion of Section \ref{sec:finite-graph-restrictions} both suggest that the dependency is not essential, even with non-empty non-maximal nodes in \(T\) the result is a decomposable graph. Moreover, non-empty non-maximal cliques are not identifiable in the decomposable graph form. Therefore, the dependence of \(\z^{\star}_{.i}\) on \(\Z^{\star}_{-(.i)}\) is only meaningful when the conditioning on the true \(T\) in a generation processes while invoking \eqref{eq:nei-bound-clique-basic}. When the true \(T\) is unknown and the junction tree \(T_{\G}\) is used instead, or when using \eqref{eq:nei-bound-clique-basic-no-cond}, such dependency is obsolete.

Therefore, for an observed \(N_{v}\)-node decomposable graph \(\G\) with a connected junction tree \(T_{\G}\), its $N_c\times N_v$ biadjacency matrix \(\Z^{\star}\) has the joint distribution
{\small
  \begin{equation}\label{eq:joint-distribution-general-W-TG}
    \begin{aligned}
      \P(\Z \mid T_{\G}) &= \prod_{i=1}^{N_{v}} \P(\z_{.i} \mid T_{\G}) \\ &= \prod_{k=1}^{N_{c}}\Bigg \{ W(\vartheta'_{k(\theta')}, \vartheta_{i}) \Bigg \}^{z_{ki}}\Bigg \{ 1-  W(\vartheta'_{k(\theta')}, \vartheta_{i})\Bigg \}^{(1-z_{ki})\deltanei_{ki}},
    \end{aligned}
  \end{equation}}
  where \(\deltanei_{ki}\) now depends on \(T_{\G}\). In fact, \eqref{eq:joint-distribution-general-W-TG} shows that the choice of a junction tree only affects the joint distribution through the component \(\deltanei_{ki}\). Therefore, assuming a uniform distribution over the set of possible junction trees, the choice of \(T_{\G}\) over an alternative junction tree \(T'_{\G}\) can be made with posterior ratios as
  \[\log\Bigg \{\frac{\P(T_{\G}\mid \Z)}{\P(T'_{\G}\mid \Z)} \Bigg \}= \sum_{i=1}^{N_{c}}(1-z_{ki})(\deltanei_{ki}- \deltanei'_{ki})\log\Bigg \{ 1-  W(\vartheta'_{k(\theta')}, \vartheta_{i})\Bigg \}.\]

The following section applies \eqref{eq:joint-distribution-general-W-TG} for specific examples.
\subsection{The multiplicative model}
The {\it multiplicative model of linkage probability} encompass a wide class of network generating models, where the link probability of the \((i,j)\)-th edge has a general form of 
\begin{equation}
(i,j) \mid p_i, p_j \sim \text{Bernoulli}(p_i p_j), \quad p_i \in [0,1].
\end{equation}

Examples of such  models are \citet{bickel2009nonparametric,chung2002connected} and \citet{olhede2012degree}. A multiplicative form of the function \(W\) can be defined as 
\begin{equation} \label{eq:separable-W}
W(x,y) = f(x)f(y), \quad x,y \in \Rp,\quad \text{for an integrable } f: \Rp \mapsto [0,1].
\end{equation}

The marginals are also functions of \(f\) as \(\overline W_1(s)=\overline W_2(s) = af(s)\) where \(a = \int_{\Rp}f(x)\d x\). 
A natural choice for \(f\) is a continuous density function, where \(a=1\), more generally, a cumulative distribution function (CDF) or the complementary (tail distribution) CDF can also be used. 

\begin{figure}[!ht]
  \centering
\includegraphics[width=0.3\textwidth]{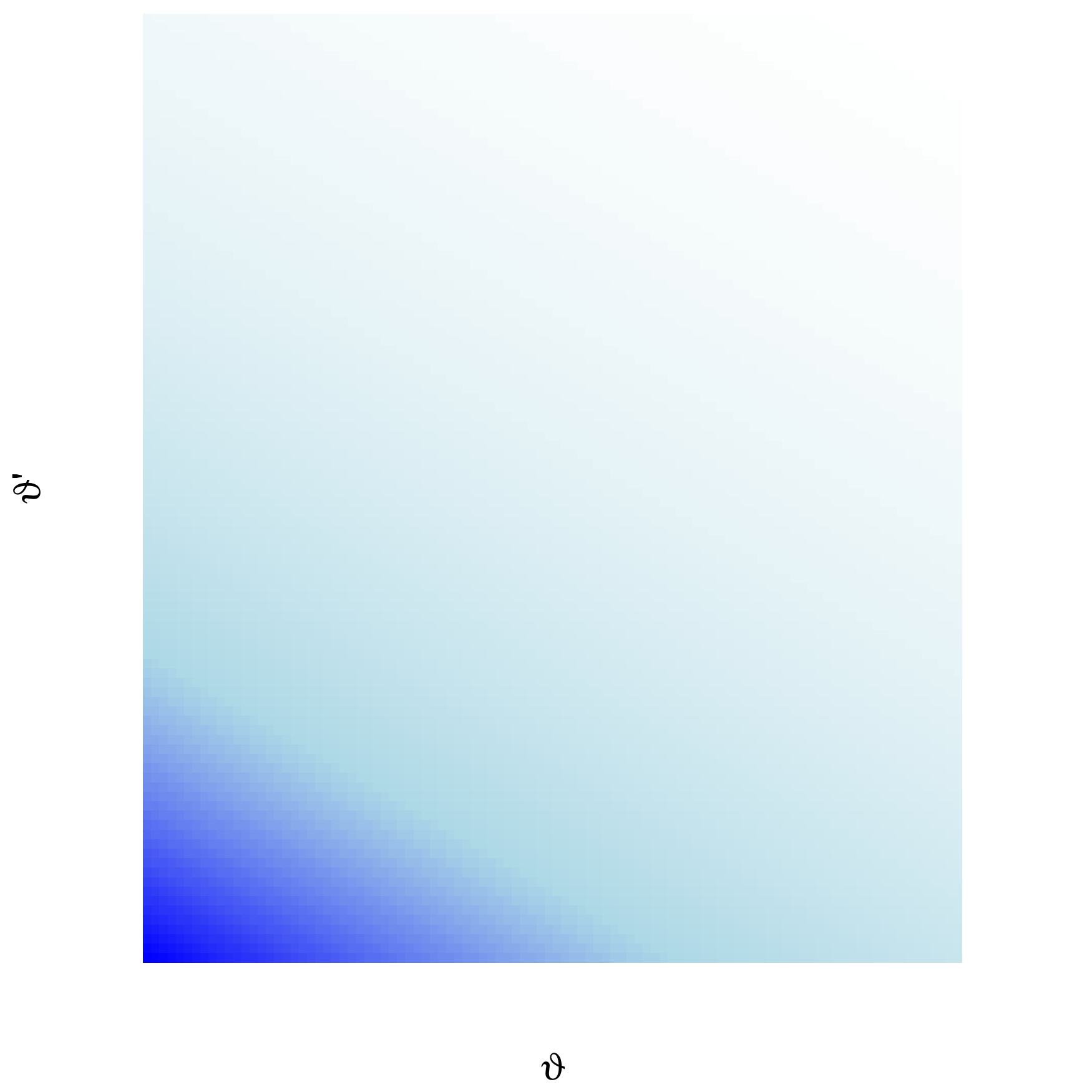}
\caption{Density of $W(x,y) = \exp(-(x + y))$.}
\label{fig:ind-marginals-tree-heatmap}
 \end{figure}

\begin{example}[Tail of an exponential distribution, fast decay]\label{exm:exponential-survival-function}
Let \(f\) be the tail of a unit rate exponential distribution, as \(f(x) = \exp(-\lambda x)\), such that 
\begin{equation} \label{eq:exponential-W-density}
  W(x,y) = e^{-\lambda(x+y)}, 
\end{equation}
with Figure \ref{fig:ind-marginals-tree-heatmap} showing the density with \(\lambda=1\).
\end{example}

\begin{figure}[!ht]
  \centering
\subfloat[10-node tree]{\includegraphics[width=0.33\textwidth]{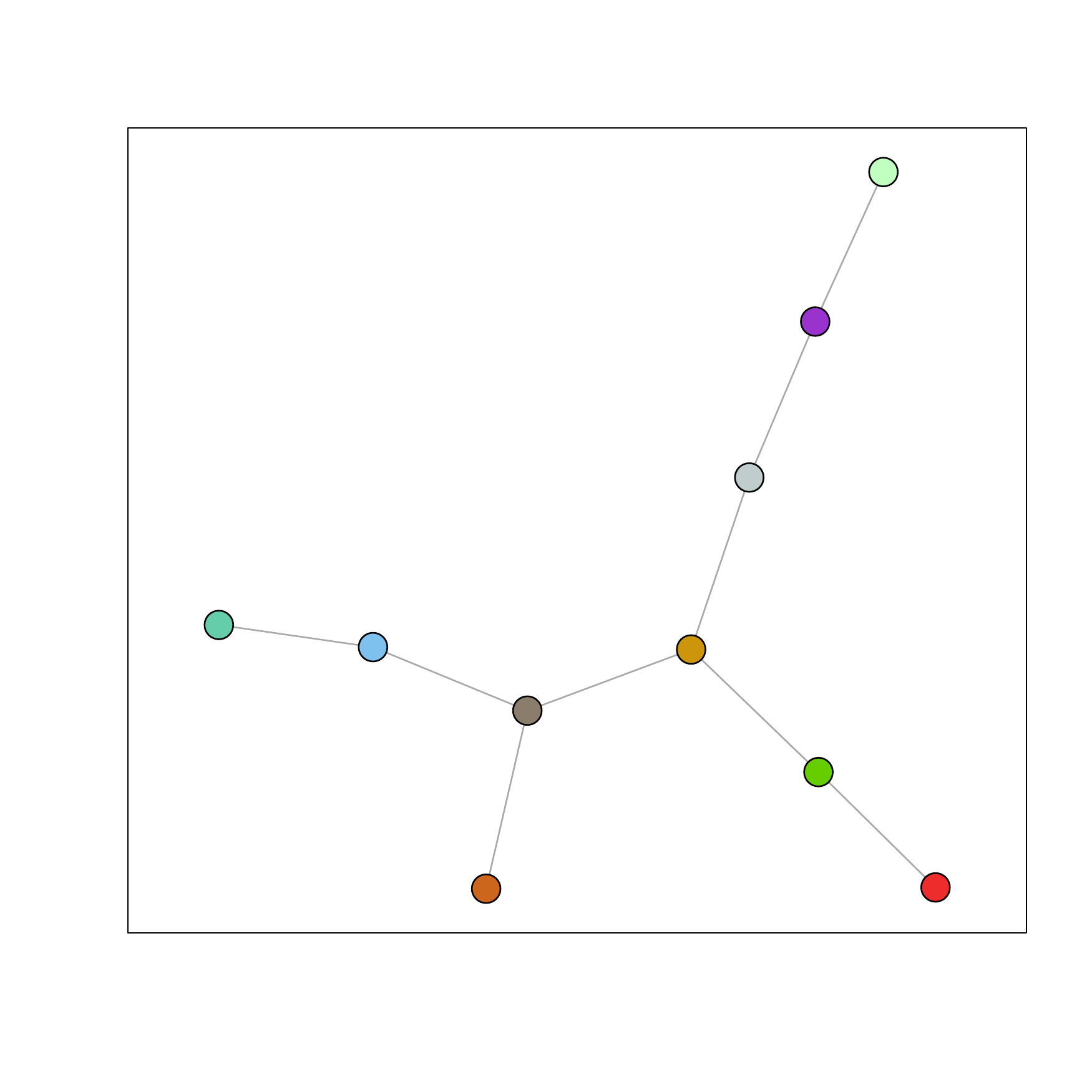}\label{fig:10-node-junction-tree}}
\subfloat[\((c=2, r=10)\)]{\includegraphics[width=0.33\textwidth]{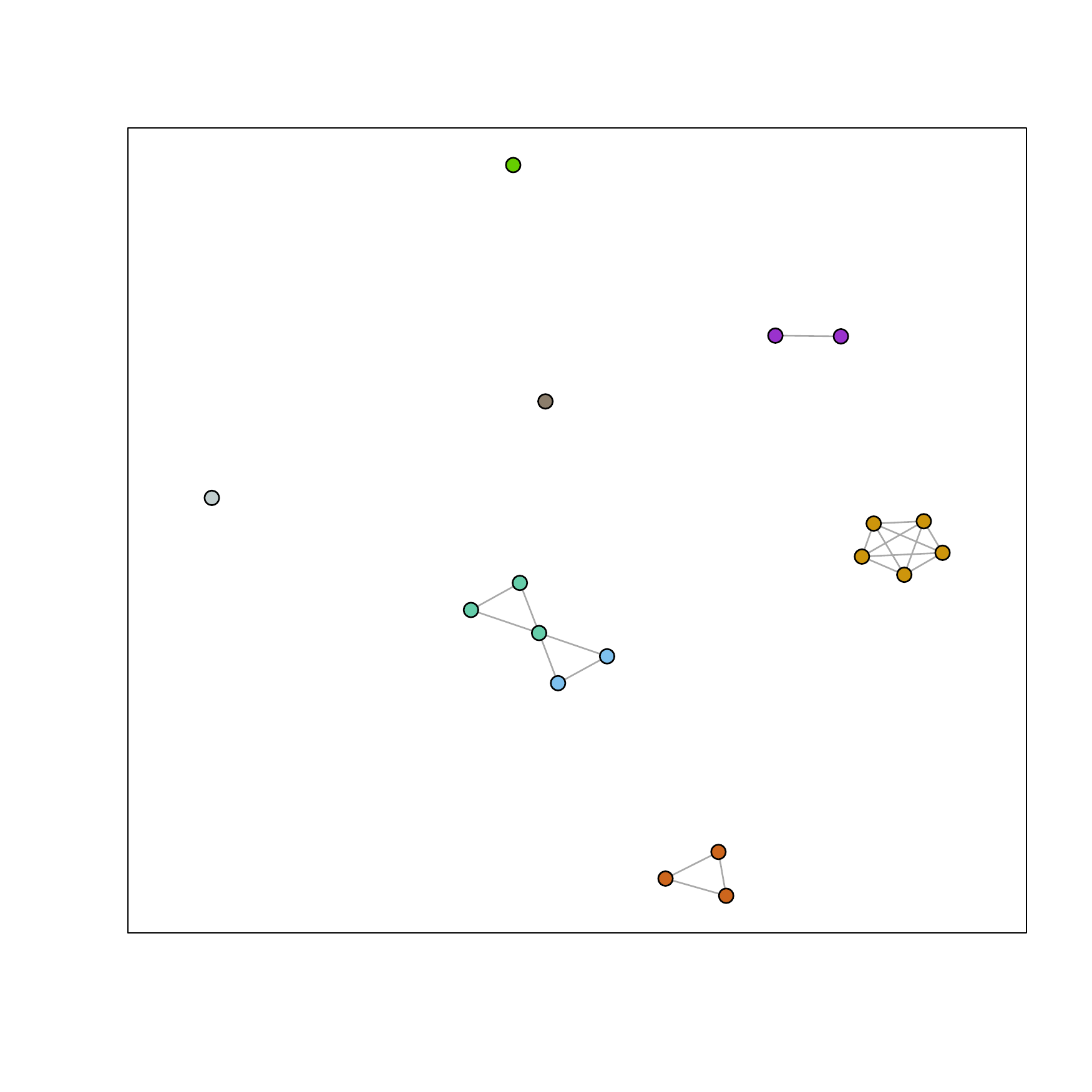}}
\subfloat[\((c=2, r=20)\)]{\includegraphics[width=0.33\textwidth]{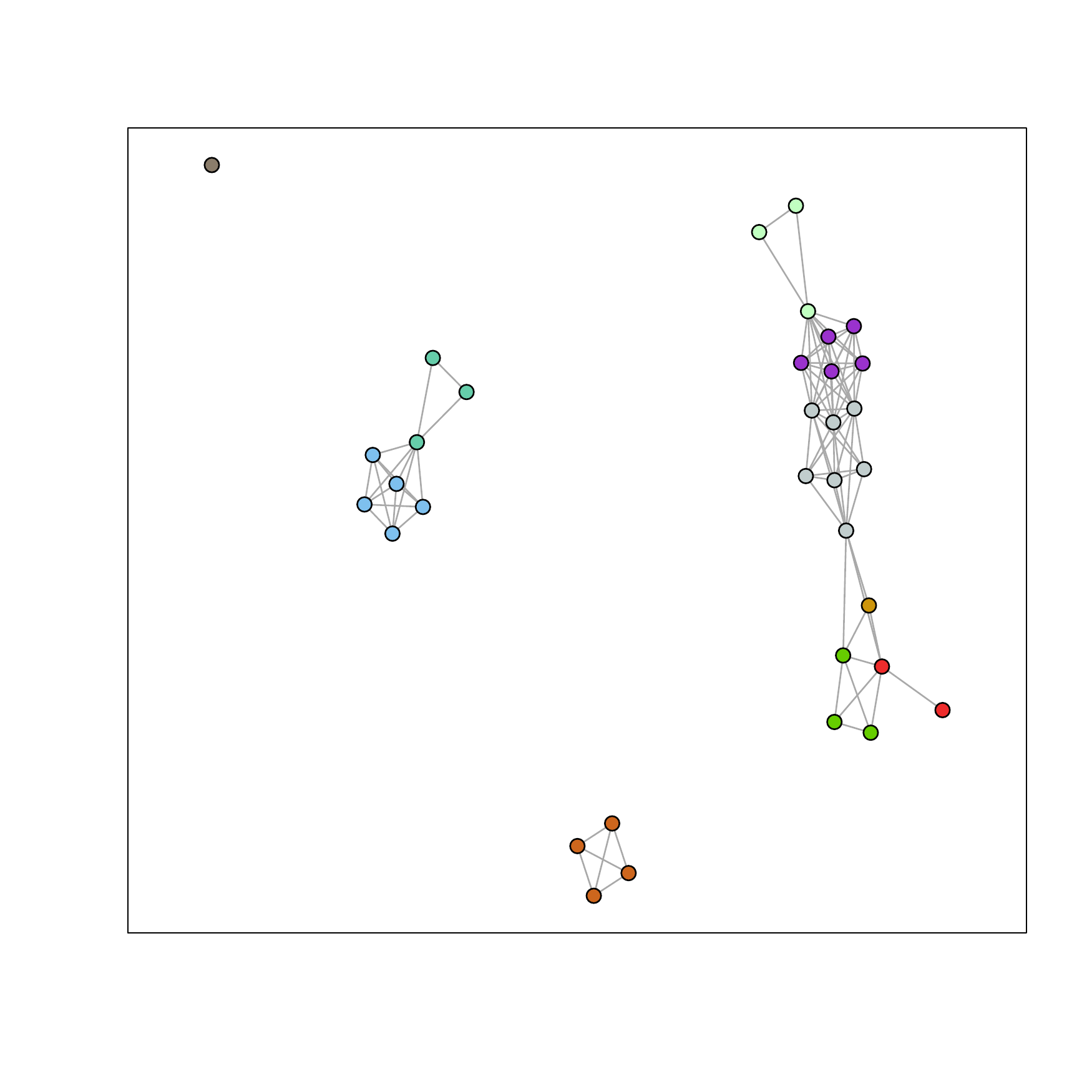}}

\vspace{-0.5cm}
\subfloat[\((c=2, r=50, \lambda=1)\)]{\includegraphics[width=0.33\textwidth]{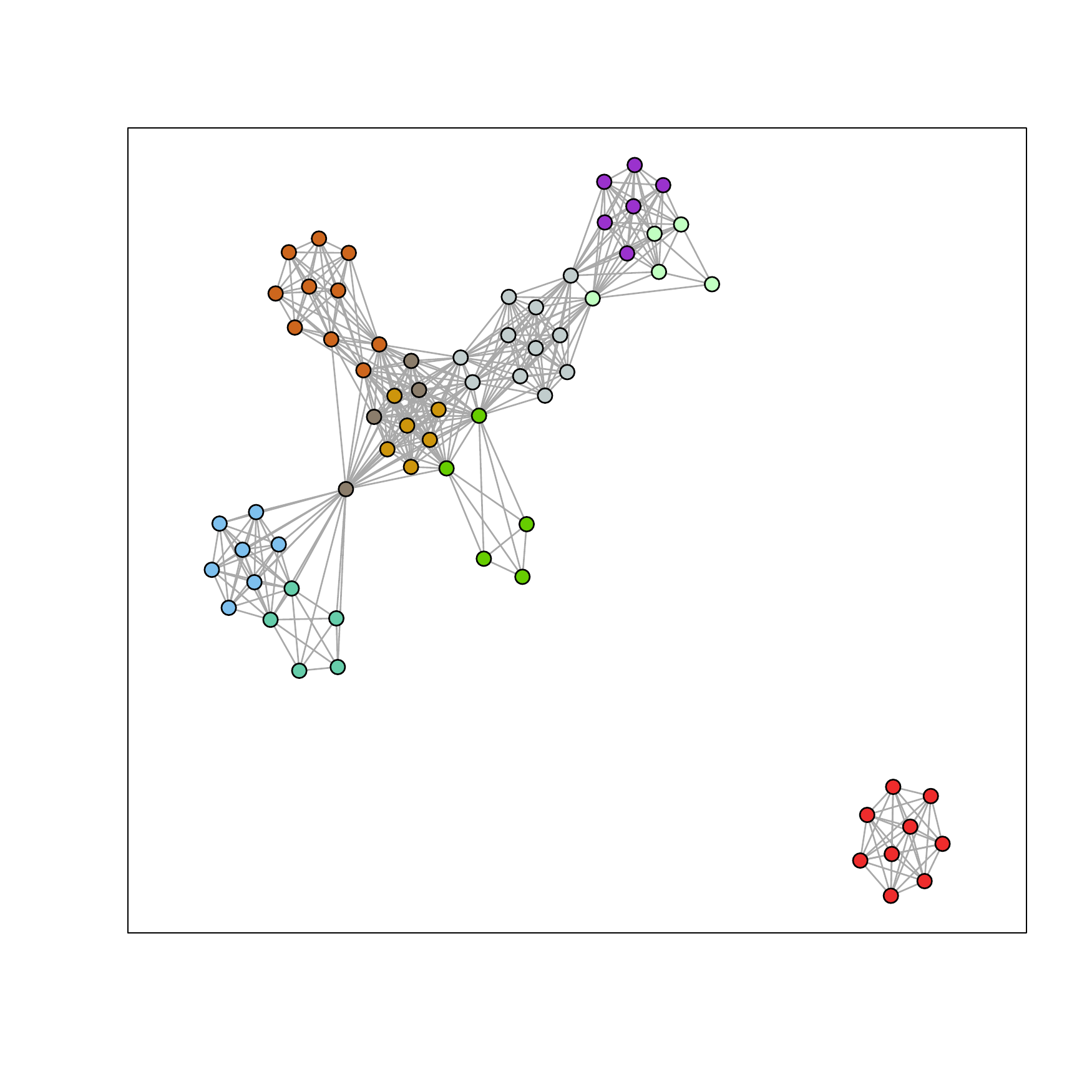}\label{fig:/Z-r50-c2-lambda1}}
\subfloat[and \(\lambda =5\)]{\includegraphics[width=0.33\textwidth]{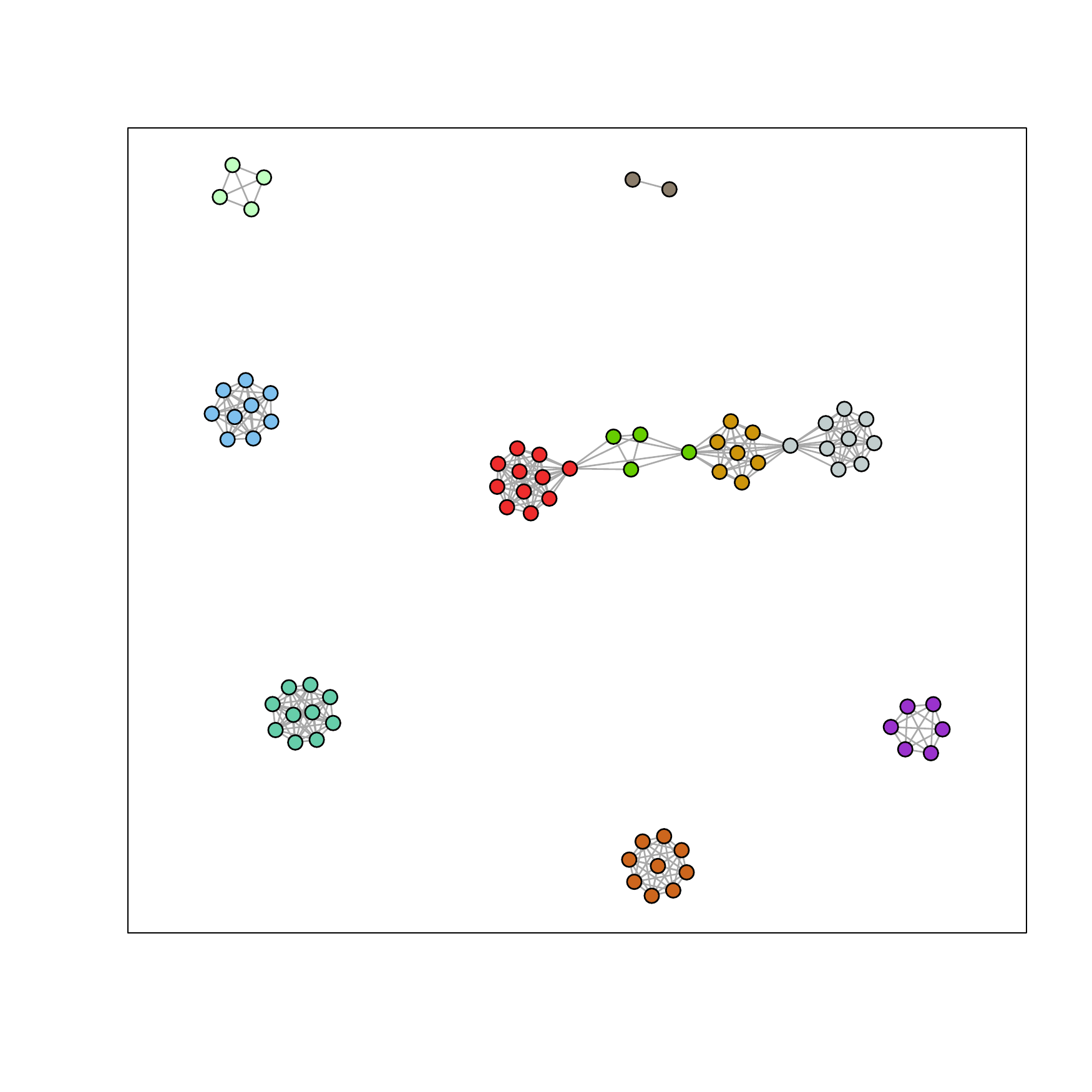}\label{fig:/Z-r50-c2-lambda5}}
\subfloat[and \(\lambda =1/5\)]{\includegraphics[width=0.33\textwidth]{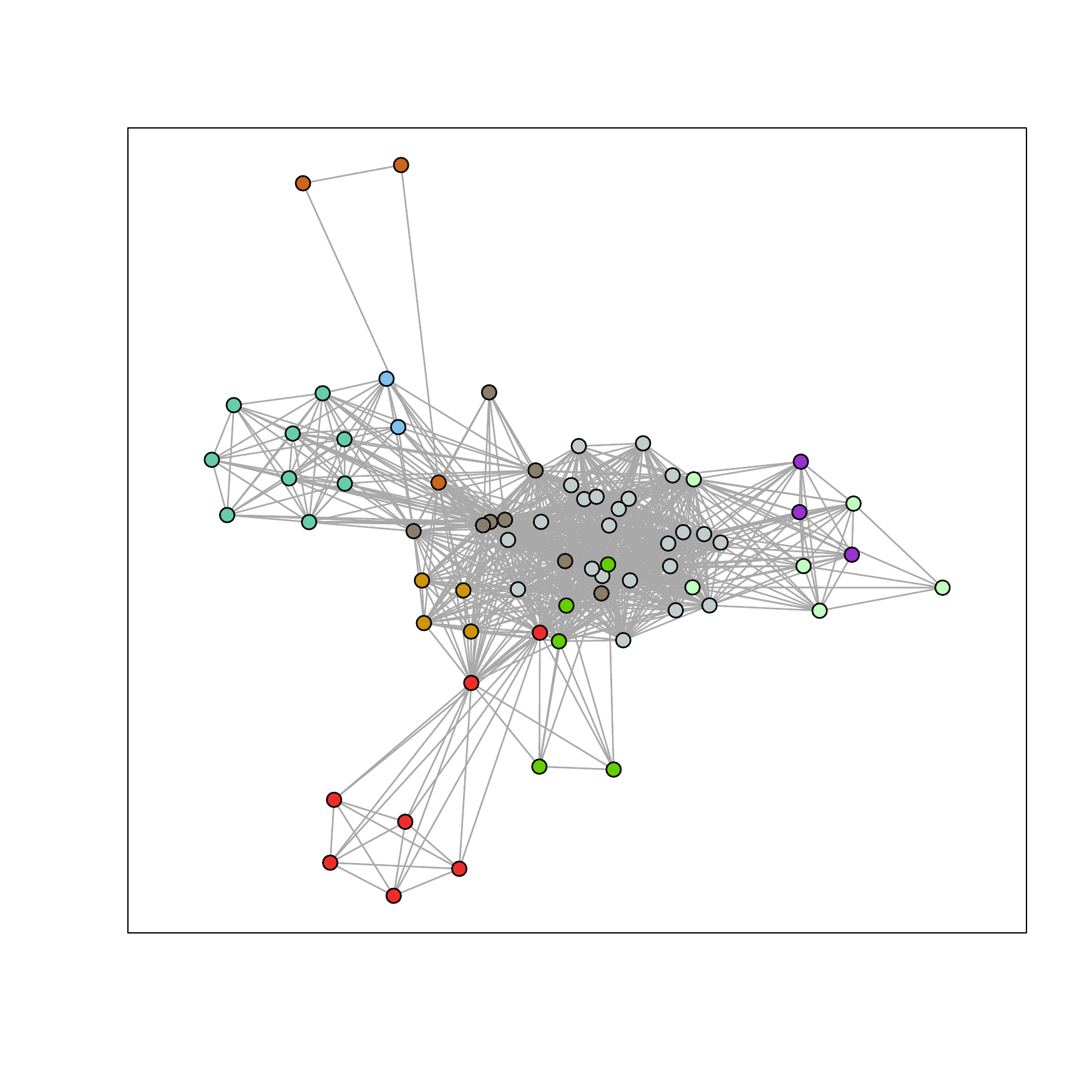}\label{fig:/Z-r50-c2-lambda2}}

\vspace{-0.5cm}
\subfloat[\((c=2, r=50)\)]{\includegraphics[width=0.33\textwidth]{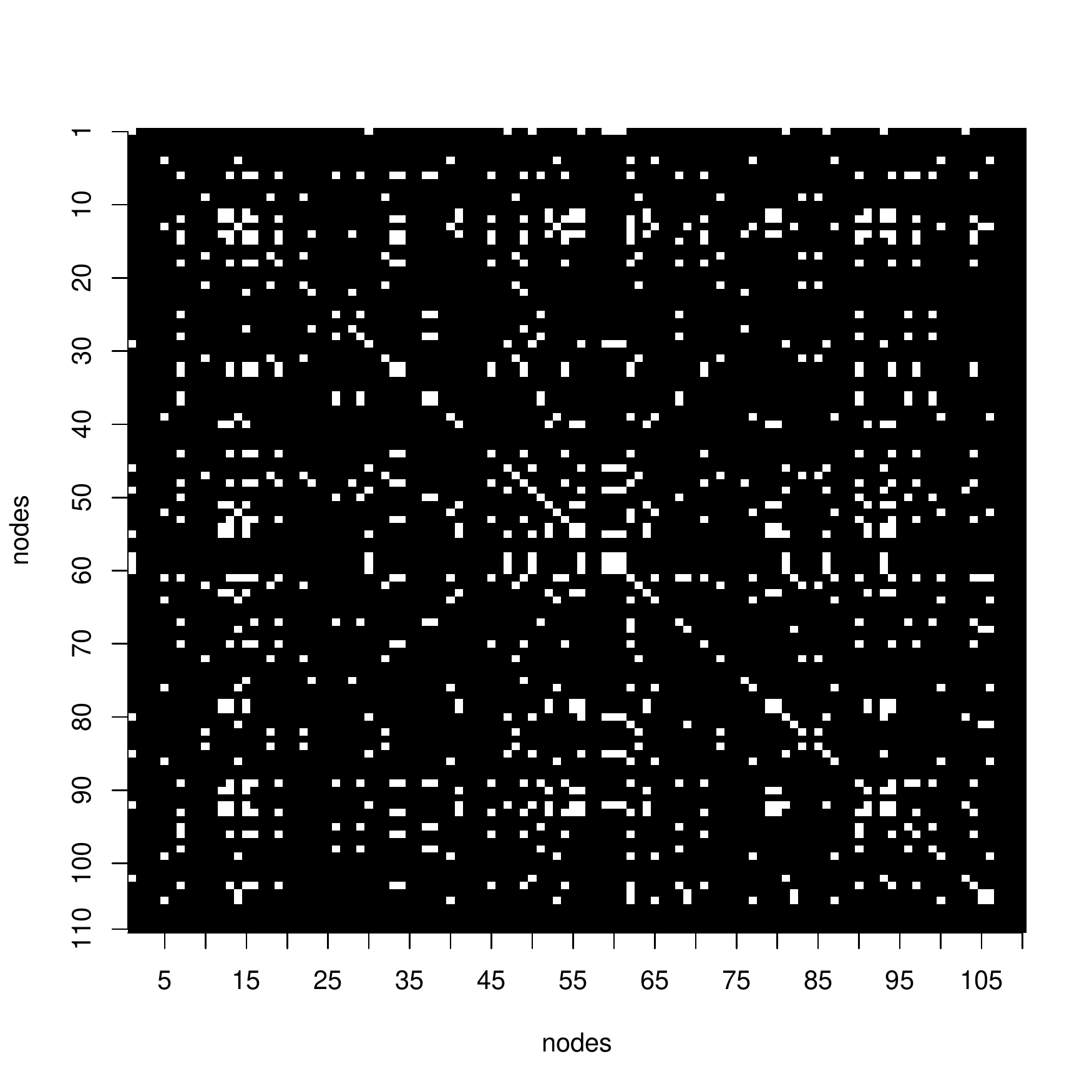}}
\subfloat[and \(\lambda =5\)]{\includegraphics[width=0.33\textwidth]{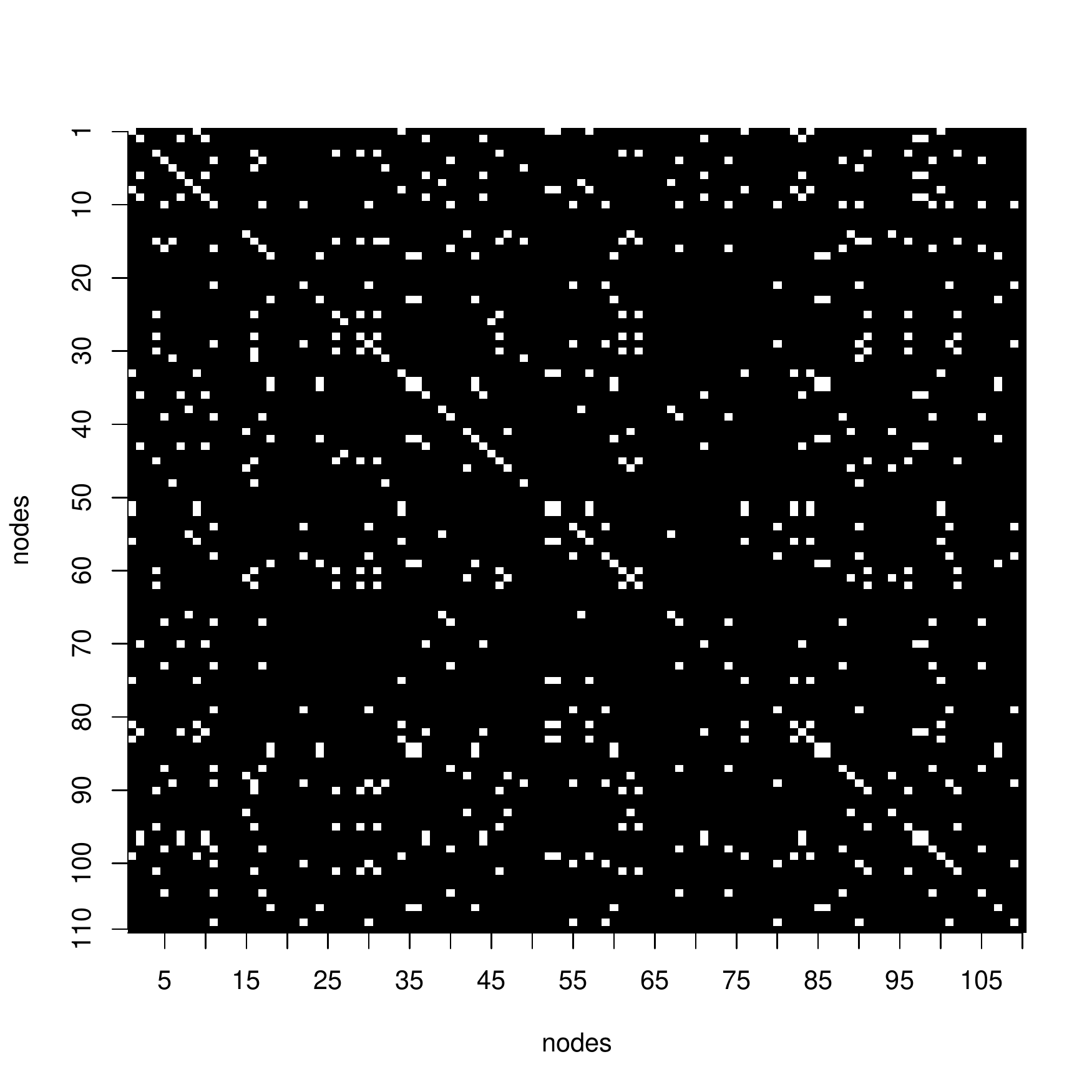}}
\subfloat[and \(\lambda =1/5\)]{\includegraphics[width=0.33\textwidth]{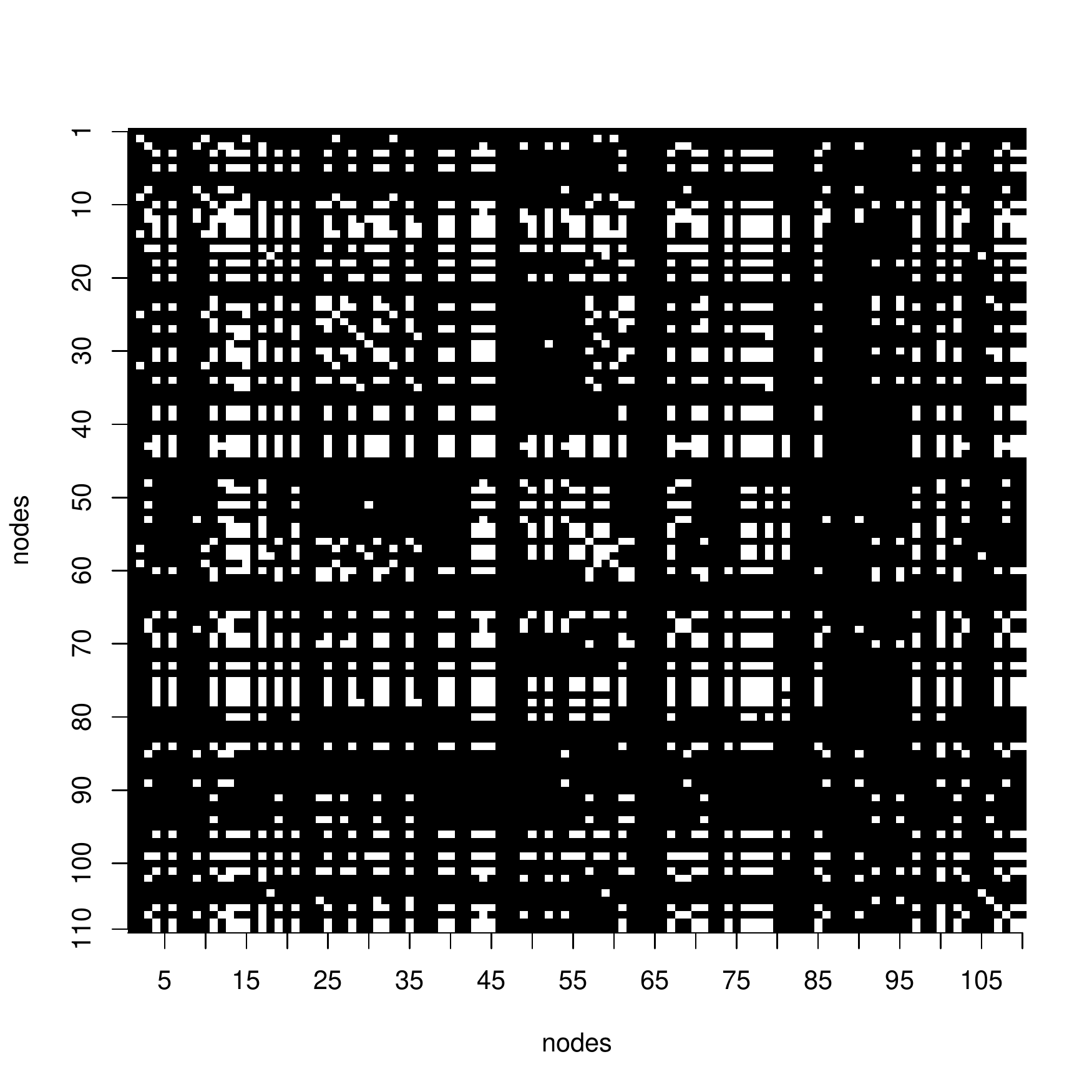}}
\caption{Different size realizations from $W(x,y) = \exp(-\lambda(x + y))$ with a fixed the 10-node tree (top left). The top and middle panels are the decomposable graphs resulting from different size realization settings, the middle panel illustrates the effect of varying $\lambda$ for the same parameter set $\{(\theta_i, \vartheta_i)\}$ generated from a $(c=2, r=50)$-truncation, the corresponding \(\G\) adjacency matrices are in the bottom panel.}
\label{fig:realization-from-exponential-example}
 \end{figure}

 Figure \ref{fig:realization-from-exponential-example} illustrates different size realizations from \eqref{eq:exponential-W-density} for the same 10-node tree (Subfigure \ref{fig:10-node-junction-tree}), where \(\lambda=1\). Each realization in the top panel is based on a different \((c,r)\)-truncation of the node domain. The middle panel illustrates the effect of varying the scaling parameter \(\lambda\), therefore, a single node parameter set \(\{(\theta_i, \vartheta_i)\}\) is used across the panel. The bottom panel plots the adjacency matrix of the corresponding decomposable graph in the upper subfigures. It is evident, high values of \(\lambda\) separate the graph in \ref{fig:/Z-r50-c2-lambda1}, while lower values support more cohesion, \ref{fig:/Z-r50-c2-lambda2}. 
 
\begin{example}[Beta multiplicative priors] Let \(f_{i}(x) \sim \text{Beta}(\alpha_i,1)\), for \(x \in \Rp\), a multiplicative form for \(W\) with Beta kernels is
\begin{equation} \label{eq:beta-prior-1}
  W(x,y) = f_{1}(x)f_{2}(y).
\end{equation}
By the ordering of the unit rate Poisson process \((\vartheta_{i})\), a Beta random variable can be sampled as  \(\exp(-(\vartheta_{(i+1)}- \vartheta_{(i)})/\alpha) \sim \text{Beta}(\alpha,1)\). Therefore, using distributional equality, the generating sequential scheme in \eqref{eq:sequential-sampling-part2} could be equivalently used with the following modification:
{\small
  \begin{equation}\label{eq;beta-prior-2}
  \begin{aligned}
    \vartheta'_k \mid \alpha_1  \stackrel{iid}{\sim} \text{Beta}(\alpha_1, 1),&\quad \vartheta_i \mid \alpha_2 \stackrel{iid}{\sim} \text{Beta}(\alpha_2, 1), \quad W(\vartheta'_k,\vartheta_i)= \vartheta'_k\vartheta_i.
  \end{aligned}
\end{equation}}
\end{example}

\subsubsection{Posterior distribution for the special case of a single marginal}
A node-clique connection probability under a single marginal is when \(W\) of the form
\[ W(x,y)= f(x), \quad \text{or} \quad W(x,y)=f(y),\quad \text{with } f:\Rp \mapsto [0,1].\]

Under such parametrization, a posterior distribution of \(f \mid \Z^{*}, T\) is easily accessible through \eqref{eq:joint-distribution-general-W} and \eqref{eq:joint-distribution-general-W-TG}, where \(\Z^{*}\) is a mapped realization from a decomposable graph \(\G\). Moreover, when \(f(\vartheta_{i})=f_{i} \sim \) Beta(\(\alpha, 1\)), as in Example \ref{eq:beta-prior-1}, the posterior distribution of \({f_{i}} \mid \Z^{*}, T_{\G}\) is 
\begin{equation}
 f_i \mid \Z^{*}, T_{\G} \sim \textsf{Beta}(\alpha +m_i, 1+ m_i^\deltanei ).
\end{equation}
where \(m_i = \sum_{k=1}^{N_c} z_{ki}\) and \( m_i^{\deltanei} =\sum_{k=1}^{N_c} \deltanei_{ki}\). The marginal joint distribution is
\begin{equation}
 \P(\Z^{*} \mid T_{\G}) = \alpha^{N_v}\prod_{i=1}^{N_v} \frac{\Gamma(\alpha +
  m_i)\Gamma(m_i^\deltanei + 1)}{\Gamma(\alpha + m_i + m_i^\deltanei + 1)} .
\end{equation}

\begin{figure}[ht!]
  \centering
\subfloat[20-node tree]{\includegraphics[width=0.3\textwidth]{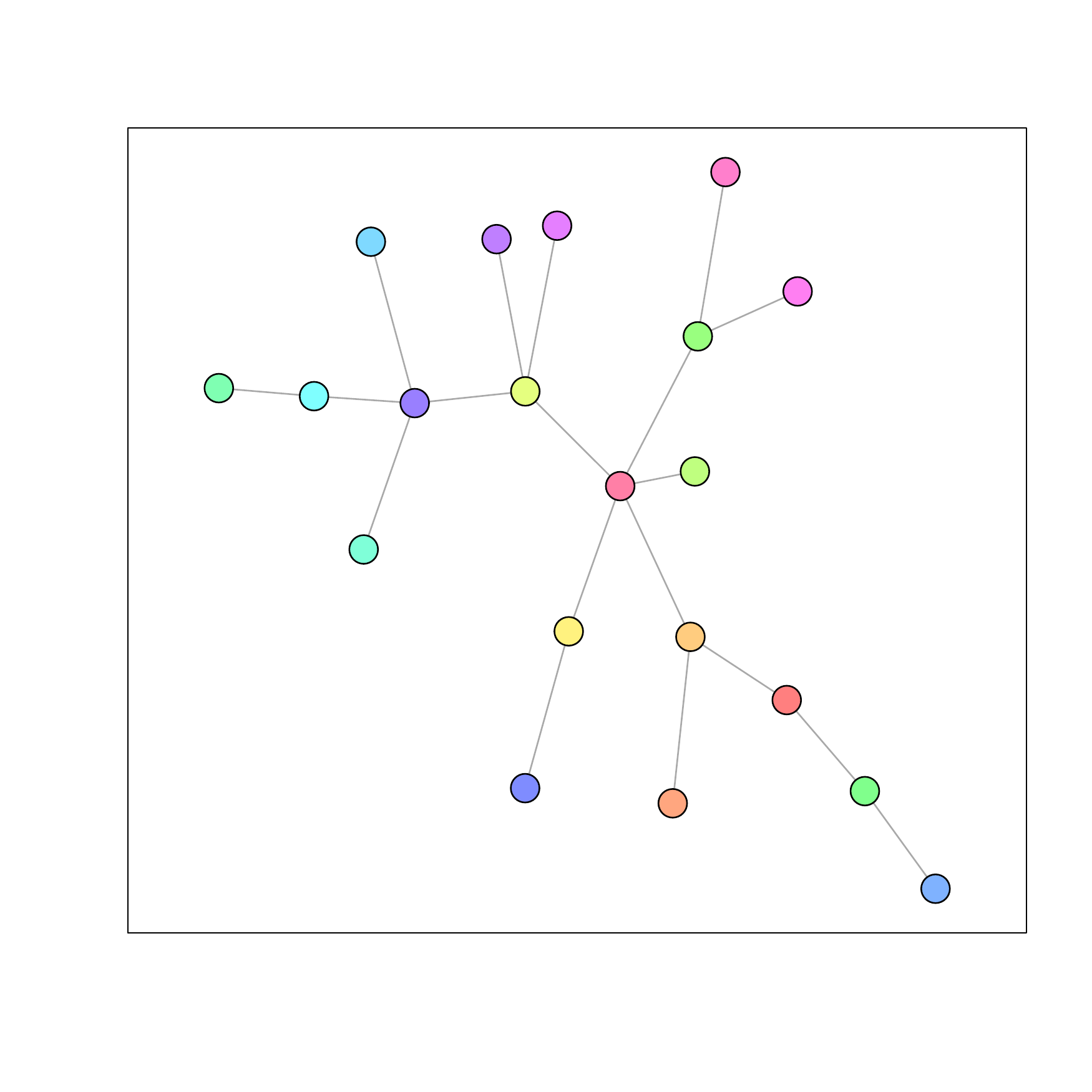}}\quad\subfloat[103-node \(\G\)]{\includegraphics[width=0.3\textwidth]{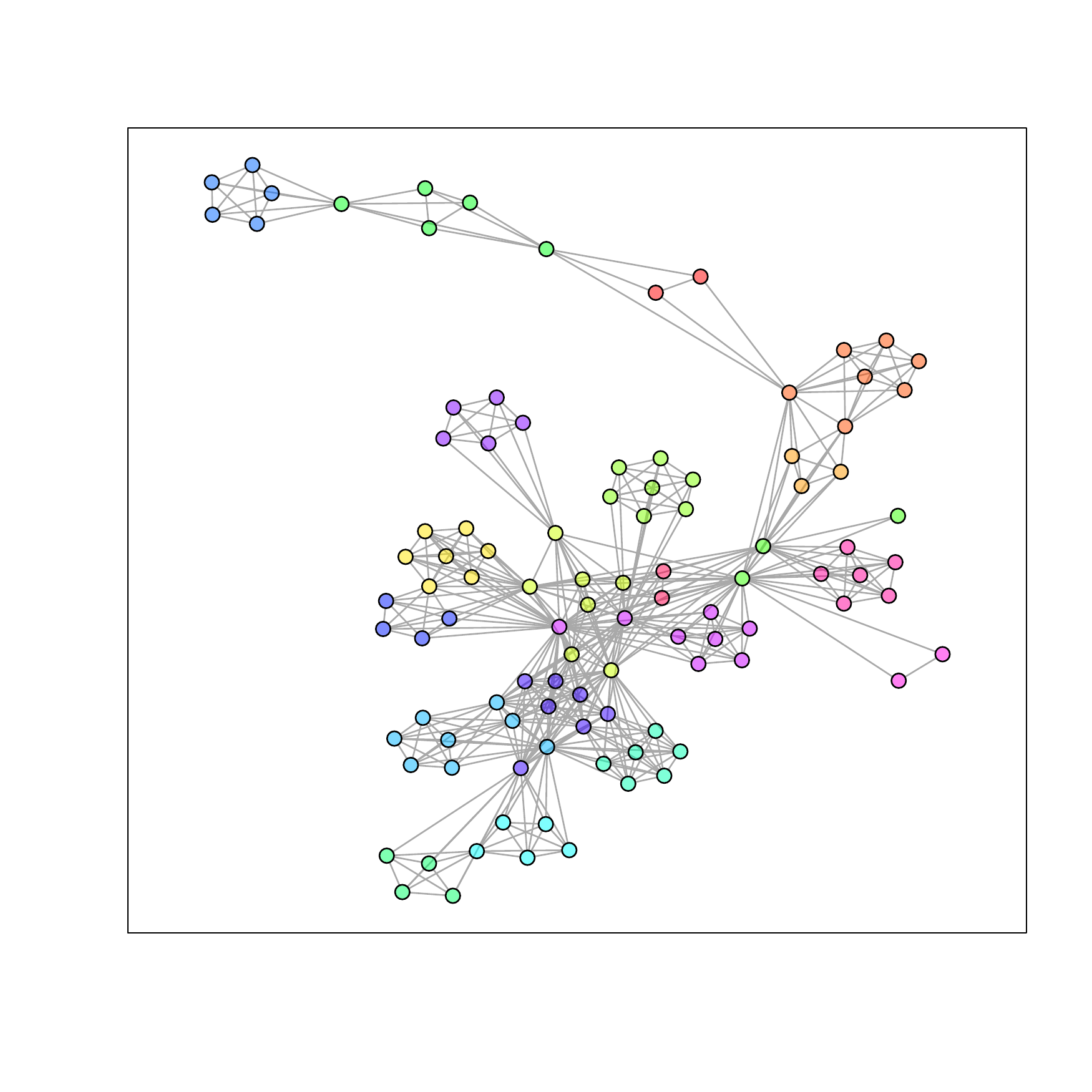}}\quad\subfloat[trace plot of \(\vartheta\)]{\includegraphics[width=0.3\textwidth]{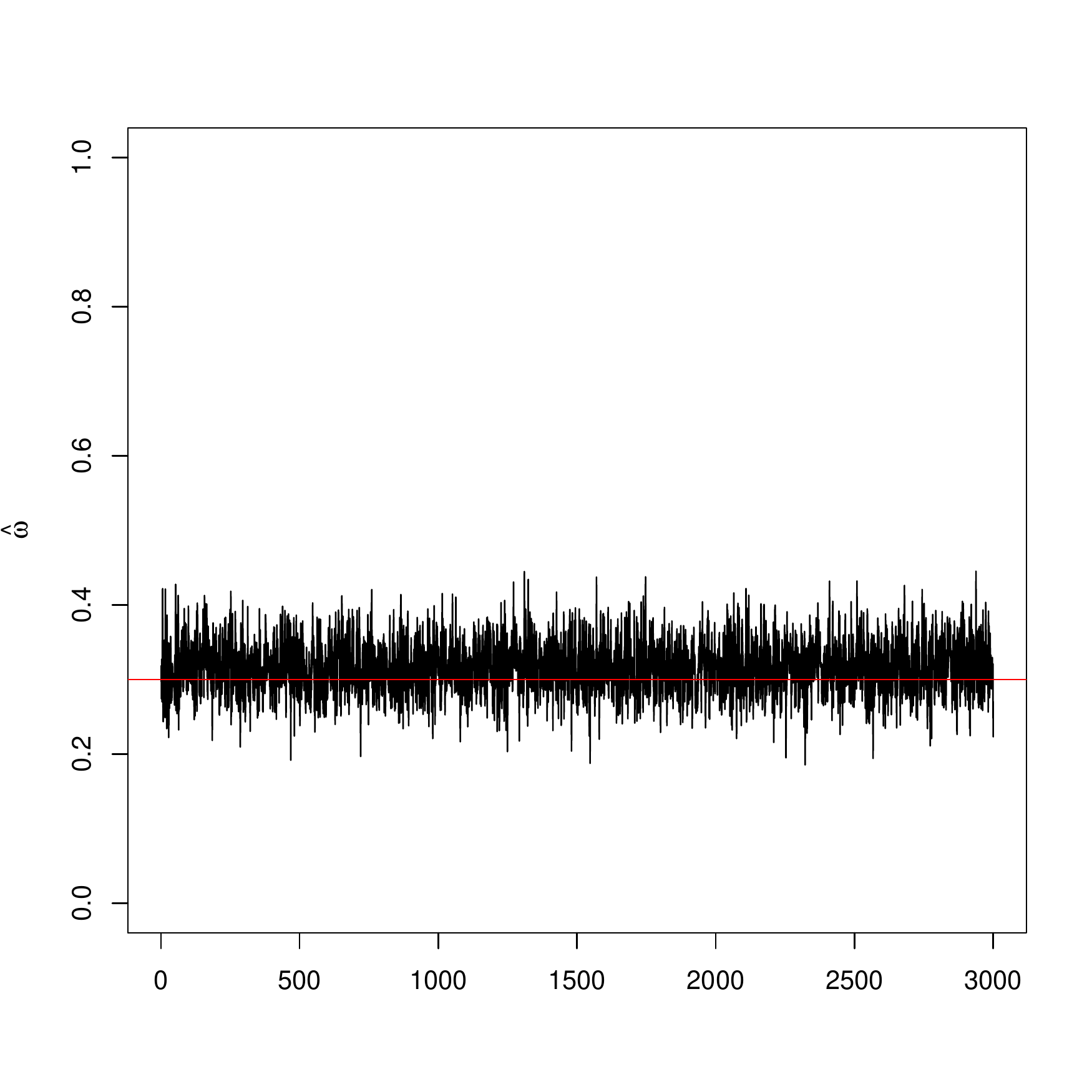}}
  \caption{Tree, decomposable graph, and the posterior MCMC trace plot of $\vartheta_i = \vartheta = 0.3$, for the case $W(\vartheta'_k, \vartheta_i)= \vartheta$.}
\label{fig:post_uni}
\end{figure}

Figure \ref{fig:post_uni} shows the posterior distribution when \(\vartheta_i = \vartheta\;\; \forall i\), for a Beta prior, for a decomposable graph of 20 cliques and 103 nodes.

A joint conditional distribution can be achieved for the case when both marginals are used. Nonetheless, the product form in  \eqref{eq:joint-distribution-general-W-TG} does not grant an easy access to the posteriors. Section \ref{sec:cox-processes} introduces an alternative parametrization that transforms the product to a sum in the log-scale, thus allowing a direct access to the posteriors.

\subsection{The log transformed multiplicative model}\label{sec:cox-processes}
Many models for random graphs parameterize the probability of an edge through a multiplicative form in the logarithmic scale. For example, the work of 
\cite{caron2014sparse,CaronBipartite} and few examples in \cite{veitch2015class}. Under such parameterization, the form of \(W\) is
\begin{equation}\label{eq:log-multiplicative} W(x,y) = 1-\exp \big (-xy\big ), \quad x,y \in \Rp.
\end{equation}

The form in \eqref{eq:log-multiplicative} is generally referred to as the Cox processes, since it can be seen as the probability of at least one event of a Poisson random variable having a mean measure as a unit rate Poisson process, hence a doubly stochastic distribution.

\subsubsection{Posterior distribution for the two marginals}
Let \(\G\) be an observed \(N_{v}\)-node decomposable graph with a connected junction tree \(T_{\G}\) of \(N_{c}\) maximal cliques. Let \(\Z^{*}\) be its $N_c\times N_v$ biadjacency matrix, with no empty rows or columns. According to \eqref{eq:joint-distribution-general-W-TG}, the joint conditional distribution of \(\Z^{*} \mid T_{\G} \) is
{\small
  \begin{equation}\label{eq:cond-dist-theta-theta}
  \P(\Z^{*} \mid T_{\G}) = \prod_{k=1}^{N_c}\prod_{i=1}^{N_v}
  \bigg( 1-\exp(-\vartheta'_{k} \vartheta_i)\bigg )^{z^{*}_{ki}}\exp \bigg (-\vartheta'_{k} \vartheta_i(1-z^{*}_{ki})\deltanei_{ki}\bigg )
\end{equation}}
where \(\deltanei_{{ki}}\) as in \eqref{eq:posterior-dist-delta-quantity}.

The product form in \eqref{eq:cond-dist-theta-theta} does not grant simple posterior expressions. By introducing an intermediary latent variable, as a computational trick, one can transform the product of densities to a sum in the exponential scale, in a manner similar to the Swendsen-Wang algorithm \parencite{Swendsen87}. Reparameterize \(z^{*}_{ki}\) using a latent variable \(s_{ki}>0\) such that
\begin{equation}
  z^{*}_{ki} = 
  \begin{cases} 1 & s_{ki}<1 \\
    0 & s_{ki}=1. 
  \end{cases}
\end{equation}

In a sense that \(z^{*}_{ki}\) is completely determined by \(s_{ki}\). The conditional model becomes
{\small
\begin{equation}
\P(z^{*}_{ki} = 1 \mid \Z^{*}_{-(ki)}, T_{\G}) = \E[\chi_{(0,1)} \mid \Z^{{*}}_{-(ki}, T_{\G}] = \P(s_{ki}<1 \mid S_{-(ki)}, T_{\G}), 
\end{equation}
}
where \(\chi_{(a,b)}\) is an atomic measure with a value of 1 on the interval \((a,b)\), otherwise 0. Note that \(S\) carries the same probability events as \(\Z^{*}\). A possible choice for the distribution of \(s_{ki} \mid S_{-(ki)}\) is a 1-inflated exponential distribution as
\begin{equation}
  p(s_{ki} \mid S_{-(ki)}) = \tau_{ki}\exp(-\tau_{ki}s_{ki})\chi_{(0,1)}(s)+ \exp(-\tau_{ki})\chi_{1}(s), 
\end{equation}
for a parameter \(\tau_{ki}\) and an atomic measure at 1, \(\chi_{1}\). The conditional joint density of \((z^{*}_{ki}, s_{ki})\), given \(\theta'_{k} \in \Tbd{}{i} \cup \Tn{}{i}\), is
{\small
 \begin{equation}
  p(z^{*}_{ki}, s_{ki}\mid T_{\G}) =\vartheta'_{k} \vartheta_{i}\exp\big (-\vartheta'_{k} \vartheta_{i}s_{ki} \big) \chi_{(0,1)}(s)  + \exp\big ( -\vartheta'_{k} \vartheta_{i} \big)\chi_{1}(s).
\end{equation}
}

It is straightforward to show that the joint conditional distribution of \((\Z^{*}, S)\) is
{\small
\begin{equation} \label{eq:cond-dist-theta-theta-extra-var}
  \begin{aligned}
    \P(\Z^{*},S \mid,T_{\G}) 
  &= \bigg [\prod_{k=1}^{N_{c}} {\vartheta'_{k}}^{m_{k}} \bigg ] \bigg [\prod_{i=1}^{N_{v}} \vartheta_{i}^{n_{i}} \bigg ] \exp \bigg (-\sum_{k, i}^{N_{c}, N_{v}}\vartheta'_{k}\vartheta_{i}s_{ki}(z^{*}_{ki} + \deltanei_{ki})\bigg),
\end{aligned}
\end{equation}}
where \(m_{k} =\sum_{i=1}^{N_{v}}z^{*}_{ki}\) and \(n_{i} = \sum_{k=1}^{N_{c}} z^{*}_{ki}\)

Under \eqref{eq:cond-dist-theta-theta-extra-var}, the posterior distribution for the affinity parameters are
{\small
\begin{equation}
  \begin{aligned}\label{eq:posterior-dist-two-marginals}
    \P(\vartheta'_{k} \mid \Z^{*}, S, T_{\G}) &\propto {\vartheta'_{k}}^{m_{k}}\exp\bigg ( - \vartheta'_{k} \sum_{i=1}^{N_{v}} \vartheta_{i} s_{ki}(1-z^{*}_{ki})\deltanei_{ki}\bigg)\omega(\vartheta'_{k}),\\
    \P(\vartheta_{i} \mid \Z^{*}, S, T_{\G}) &\propto {\vartheta_{i}}^{n_{i}}\;\,\exp\bigg ( - \vartheta_{i} \sum_{k=1}^{N_{c}} \vartheta'_{k} s_{ki}(1-z^{*}_{ki})\deltanei_{ki}\bigg)\omega(\vartheta_{i}),
  \end{aligned}
\end{equation}}
where \(\omega\) is a prior distribution. A natural conjugate prior for \eqref{eq:posterior-dist-two-marginals} is the Gamma distribution. Conditionally updating \(s_{ki}\) can be done by a truncated exponential distribution at 1 as
\begin{equation}
  s_{hj} \mid \Z^{*}, T_{\G} \sim 
  \begin{cases}
    \chi_1 \quad &  \text{if } z^{*}_{hj} =0 \\
    \text{tExp}\bigg (\vartheta'_{k}\vartheta_{i},1\bigg )\quad  &  \text{if } z^{*}_{hj} =1, \\
  \end{cases}
\end{equation}
where tExp(\(\lambda, x\)) is the exponential distribution with parameter \(\lambda\) and truncated at \(x\).

\section*{Discussion}
This work was inspired by the need for easy update rules for decomposable graphs, specially in high dimensional settings. Instead of modelling the adjacency matrix of a decomposable graph \(\G\), this work adopts a different approach by modelling a semi-latent process \(\Z\), representing the interaction between the decomposable graph nodes and latent communities
described in a tree form \(T\). We term this process as the tree-dependent bipartite graph. Like many other models for decomposable graphs, this model is iterative; updating \(\Z \mid T\) and iteratively \(T \mid \Z\).

This work investigates two possible interpretation of \(\Z\) with respect to \(\G\): i) \(\G\) is a surjective projection from \(\Z\); ii) \(\G\) is a bijective projection from \(\Z\). In ii), not all communities in \(\Z\) represent maximal cliques in \(\G\), while in i), every community in \(\Z\) is assumed to represent a maximal clique of \(\G\). So far, most known models of decomposable graphs follow the relation in ii).

The proposed framework has several benefits, most importantly, it enables a fast sampling algorithm even for very large graphs. This is achieved by the simplicity of the Markov update conditions in \(\Z \mid T\), which also breaks the dependency among nodes, hence, decoupling the generative Markov chain into parallel chains, one for each node.

Another appealing benefit of this framework is the easy access to the set of maximal cliques, and consequently a junction tree representation. In addition, one can explicitly define the expected number of cliques per node, conditionally on a class of \(d\)-regular junction trees.

This work builds on a newly proposed framework for random graphs, in particularly the Kallenberg representation of graphs as continuous point processes. Nonetheless, in the Kallenberg representation, \(W\) is seen as a random function converging to some probability law in the limiting sense, and the objective is to estimate latter. In the random graphs settings, recent work have shown promising developments, for example \cite{castillo2017uniform,maugis2017statistical,veitch2016sampling,borgs2015private, wolfe2013nonparametric}. However, This is yet to be investigated in the decomposable graphs settings. At the current stage, this work treats \(W\) as a fixed function of known distributions, avoiding the complexities with estimating bivariate distributional functions. An approach similar to \cite{caron2014sparse}, where \(W\) is a function of completely random measures, can easily be implemented. 

\bibliographystyle{chicago}  
\bibliography{references}
\end{document}